# Barriers to Integration: Physical Boundaries and the Spatial Structure of Residential Segregation*

Elizabeth Roberto
*Princeton University*

Jackelyn Hwang
*Stanford University*

September 2017

## Abstract

Despite modest declines in residential segregation levels since the Civil Rights Era, segregation remains a defining feature of the U.S. landscape. This study highlights the importance of considering physical barriers—features of the urban environment that disconnect locations—when measuring segregation. We use population and geographic data for 20 U.S. Rustbelt cities from the 2010 decennial census and a novel approach for measuring and analyzing segregation that incorporates the connectivity of roads and the excess distance imposed by physical barriers, such as highways, railroad tracks, and dead-end streets. We find that physical barriers divide urban space in ways that reinforce or exacerbate segregation, but there is substantial variation in the extent to which they increase segregation both within and across these cities and for different ethnoracial groups. By uncovering a new source of variation in the segregation experienced by city residents, the findings have implications for understanding the mechanisms that contribute to the persistence of segregation and the consequences of segregation.

---

* Direct correspondence to Elizabeth Roberto, Department of Sociology, Princeton University, 107 Wallace Hall, Princeton, NJ 08544; Email: eroberto@princeton.edu. The authors thank Julia Adams, Richard Breen, Jacob Faber, Scott Page, Andrew Papachristos, and Jacob Rugh for helpful feedback. This research was supported by the James S. McDonnell Foundation Postdoctoral Fellowship Award in Studying Complex Systems; the Princeton Institute for Computational Science and Engineering (PICSciE) and the Office of Information Technology's High Performance Computing Center at Princeton University; and the Eunice Kennedy Shriver National Institute of Child Health & Human Development of the National Institutes of Health (NIH; Grant No. T32HD007163).

**Barriers to Integration:**
**Physical Boundaries and the Spatial Structure of Residential Segregation**

A long line of research shows that residential segregation by race—the extent to which racial groups reside in distinct places—plays an important role in perpetuating racial stratification in the U.S. (Massey 2016). In the U.S., different race groups live, on average, in distinct social environments (Logan et al. 2011), which can lead to unequal outcomes for families and individual life chances (Jargowsky 1997; Sampson 2012). Despite the decline of residential segregation levels since the Civil Rights Era, when legislation made explicit discrimination in real estate and lending illegal, segregation levels remain high, and segregation continues to be a defining feature of the U.S. landscape (Rugh and Massey 2014).

Research focuses on three primary causes for explaining the persistence of segregation: 1) socioeconomic differences, 2) residential preferences, and 3) housing market discrimination (Charles 2003). Although these factors contribute to explaining why residents of different groups tend to live in separate areas from each other, we know less about the processes that lead people to end up living where they live (Crowder and Krysan 2016). We argue that physical barriers—features of the built environment that structure how urban spaces are (dis)connected (e.g., streets and walking paths cut off by dead-ends, cul-de-sacs, walls around gated communities, highways, rivers, and railroad tracks)—are a key mechanism that contributes to these processes that facilitate the persistence of residential segregation.

Residents may live in or move to different places due to differences in socioeconomic status, residential preferences, and housing market practices, but physical barriers structure urban space by separating areas in which residents live and to which residents move, thereby restricting integration. For example, Grannis (1998) finds that areas that are connected by road networks, rather than mere proximity, have more similar racial compositions. Physical barriers also restrict social integration and physical access (Jacobs 1961; Grannis 1998). Further, physical barriers can structure residential sorting processes by providing clear demarcations to distinguish land values between one geographic space from another, to ease the categorization of areas in the housing search process, or to market as "protection" from residents on the other sides of the boundaries (Krysan and Bader 2009). Such processes limit the expansion of minority areas and maintain the social and spatial isolation of groups, having important implications for inequality by producing socially isolated environments with distinct social and economic conditions that can lead to unequal outcomes for individuals and households (Rabin 1987; Hoxby 2008; Ananat 2011). While some physical barriers, such as rivers, are features of the natural environment, others, such as Detroit's "wailing wall" (Sugrue 1996) and development of the highway system (Rabin 1987), are products of direct efforts by institutional actors to separate areas.

A key feature of physical barriers is that they are difficult to negotiate. They are distinct from spatial boundaries like streets or landmarks that carry symbolic meaning that can structure discrimination and residential preferences and prohibit social interaction but are still negotiable (e.g., white gentrifiers moving into previously minority areas separated by a symbolic racial divide) (Anderson 1990; Lamont and Molnar 2002; Hwang 2016). In contrast, dismantling physical barriers often requires institutional action, such as city-sanctioned urban planning and design that connects streets, urban spaces, and both natural and built boundaries by adding paths or building bridges or highway underpasses or removing built divisions (Rabin 1987). Thus, these boundaries serve as a distinct mechanism that facilitates the perpetuation of segregation and are the focus of our study.

We argue that physical barriers are a powerful force that exacerbates residential segregation and contributes to its persistence. Although a number of studies implicate the importance of physical barriers as important features of the built environment that contribute to the persistence of residential segregation, only a handful of studies examine how physical barriers themselves contribute to the persistence of residential segregation. Using a novel approach to measure and analyze segregation, we systematically demonstrate in our analysis that physical barriers divide urban space in ways that increase residential segregation.[1] Our study is the first to our knowledge to systematically examine how these physical barriers influence residential segregation levels across several cities. We do this by employing a novel method developed by Roberto (2015) that builds on recent developments for measuring segregation and can incorporate physical barriers. By allowing us to integrate physical barriers in measuring segregation levels, the method permits us to systematically examine how these features influence residential segregation levels. Our findings contribute to understandings of the persistence of segregation by demonstrating an important mechanism that facilitates and maintains residential segregation.

## PHYSICAL BARRIERS AND THE PERSISTENCE OF RESIDENTIAL SEGREGATION

Although segregation between blacks and whites has declined over the last forty years, segregation levels remain high, particularly in areas that historically have had the highest levels of segregation for these groups in the country (Rugh and Massey 2014). The level of segregation for Hispanics has also increased moderately, especially in areas where their population has grown rapidly, while Asians, on the other hand, have maintained moderate levels of segregation that have changed little over this time (Rugh and Massey 2014). How physical barriers separate the areas in which residents live are important yet missing pieces from explanations of the persistence of segregation.

Spatial boundaries have long played a prominent role in distinguishing spatial areas in which residents live in urban sociology. After all, the notion of neighborhoods as organizing units of residence implies a separation of residents along spatial boundaries that separate one place from another. Early Chicago school scholars Park, Burgess, and McKenzie (1925) described neighborhoods as “natural areas,” demarcated by land uses, distinctive streets and landmarks, rivers, railroad tracks, and streetcar lines, that separated racial and ethnic groups, as well as residents with distinct socioeconomic statuses and family compositions. These “natural” boundaries of the physical environment, they argued, organized the physical separation of groups of residents. Although there are distinctive streets or landmarks that carry symbolic meaning as boundaries between residents of different groups, contributing to the persistence of residential segregation (Suttles 1972; Hunter 1974; Anderson 1990; Hwang 2016), these fluid and negotiable boundaries are distinct from strong and persistent forms of spatial boundaries like

[1] Municipal boundaries and school catchment areas are other examples of spatial boundaries that may also have similar effects but are beyond the focus of this study (see Cutler and Glaeser 1995; Bischoff 2008).

physical barriers.

Figures 1 and 2 illustrate stylized examples of how two forms of physical barriers may separate areas despite close proximity and in distinct ways from symbolic streets or landmarks. The panel on the left of Figure 1 contains a predominantly white neighborhood on the north side of "Main Street" and a predominantly black neighborhood on the south side of "Main Street." In this illustration, the white neighborhood is comprised of several cul-de-sacs. The panel on the right of Figure 1 illustrates an identical spatial configuration of racial compositions but with a distinct street pattern, where the cul-de-sacs do not exist and the minor streets are instead connected to Main Street. In both figures, the residents in the white and black neighborhoods are in close proximity to each other and are in the exact same geographic location, but these figures represent two socially distinct residential environments with distinct possibilities for integration. The white neighborhood in the left panel is much more isolated than the same neighborhood in the right panel. Although Main Street may serve as a strong symbolic divide between these two neighborhoods in the moment depicted in the figures, the possibility of integration, though perhaps temporary, such that black or white residents spill over into the adjacent area, is present.

Physical barriers like disconnected roads produced by railroad tracks can work in a similar way. Figure 2 illustrates this example. In both panels, railroad tracks run east and west across the center of the figures, separating a predominantly white neighborhood to the north and a predominantly black neighborhood to the south. In the panel on the left, only major arterial roads provide access across the railroad tracks, isolating the two neighborhoods from each other. The panel on the right, however, presents an alternative scenario where there are roads that connect across the railroad tracks every two blocks. Thus, while the railroad tracks may separate who is on which side of the tracks in this figure, the connectivity of these areas present less barriers to integration. In many areas, railroad tracks are no longer in use, yet these divisions often remain.

[Figure 1 about here.]

[Figure 2 about here.]

Existing literature points to several ways in which physical barriers can maintain residential segregation. Physical barriers block the connectivity of urban spaces that roads and/or walking paths typically connect. Jane Jacobs (1961) called attention to such disconnected areas and noted that barriers created and maintained distinct social and economic conditions and prohibited social contact by limiting pedestrian traffic across separated spaces. Such conditions, Jacobs (1961) argues, results in a self-perpetuating process that maintains divisions between spaces. Grannis' (1998) work on defining neighborhoods with tertiary streets—smaller streets conducive to pedestrian traffic—further demonstrates that the connectivity of streets and roads aligns with patterns of residential segregation to a greater degree than administratively-defined boundaries like census tracts.

In some cases, physical barriers emerged out of intentional practices to socially exclude people and reinforce segregation. Gated communities emerged as attempts to reinforce segregation as a social exclusion effort, often guised to curb crime and maintain home values (Blakely and Snyder 1997; Low 2001; Atkinson and Flint 2004). Efforts to construct fences or other types of barriers reflect similar processes. For example, in Baltimore, an "eight-foot-tall spiked fence" was constructed around a public housing project in the Hollander Ridge neighborhood in 1998 to block access to the mostly white neighborhood of Rosedale, whose residents wanted the fence to "keep out crime and keep their property values up" (Schindler 2015:1957). Blakely and Snyder (1997) document how blocked streets and cul-de-sacs were also motivated by efforts to socially exclude people and reinforce segregation, and Rabin (1987) describes discontinuous street systems purposefully constructed to only connect black neighborhoods to a major artery via a single street while limiting access and exposure to an adjacent white neighborhood. In other cases, however, physical barriers are products of the natural landscape, such as rivers or other bodies of water cross-cutting through cities, or of urban planning decisions that were not necessarily intentional strategies to segregate areas. For example, the placement of railroad tracks or large roads were often contingent on the preexisting street design and proximity and accessibility considerations.

Regardless of intention, physical barriers provide a means for differentiating urban space. Whether by street design that disconnects areas or the presence of gates, walls, or fences, physical barriers can lead racially or economically distinct groups to reside separately from each other, even if they are spatially proximate to each other (Neal 2012). With distinct social and economic conditions on each side of a physical barrier, physical barriers can structure residential sorting processes that perpetuate segregation in a recursive pattern. For example, different social and economic conditions can make one area more valuable and attractive than an adjacent area to potential residents, businesses, developers, and investors (Noonan 2005). Sugrue (1996) documents how a developer constructed Detroit's "wailing wall" during the 1940s to separate an area with minority residents to avoid redlining in the adjacent area. The physical barrier provided a demarcation separating areas that others viewed as worthy of investment. Similarly, Atkinson and Flint (2004) describe how housing developers marketed physical barriers in geographically proximate areas with varying social and economic conditions as "protection" for residents to increase the value of the area on one side of the barrier.

Physical barriers can also ease the cognitive process of deciding where to search for housing or where to invest. People often limit the area in which they will consider moving before actually searching for housing to simplify the housing search process, rarely giving all areas equal consideration (Krysan and Bader 2009). By offering discrete divisions between spatial areas with distinct conditions, physical barriers can contribute to easing the cognitive processing of areas, thereby perpetuating continued reinvestment in one area but continued disinvestment in another spatially proximate one.

At the same time, the nature of physical barriers inhibits the possibility of negotiating boundaries. While many neighborhood boundaries are often flexible and fluid, cognitive mapping studies find that residents often agree on these distinct physical features as neighborhood boundaries (Chaskin 1997; Campbell et al. 2009). Research on integrated neighborhoods or neighborhoods that experience racial change identify processes of spillover or expansion into areas through the negotiation of a spatial boundary (Anderson 1990; Hwang 2015). The gentrification documented in these studies shows how integrated blocks or areas, though often temporary, increases whites' comfort and willingness to move into areas further beyond a previously symbolic boundary. Physical barriers, on the other hand, can enable stark contrasts between proximate areas to persist, thereby limiting the possibility of integration between areas.

Despite the many in ways in which past literature has implicated physical barriers and the disconnectedness that they provide in perpetuating residential segregation, only a few studies have systematically examined the relationship between physical barriers and residential segregation. Studies identifying the causal effect of segregation on some outcome using instrumental variable approaches—a statistical method requiring a variable (the “instrument”) that is correlated with the segregation measure but is unrelated to the outcome variable—have used natural boundaries (e.g., rivers) or built boundaries (e.g., railroad tracks) as instruments for their analyses (e.g., Hoxby 2000; Ananat 2011). While this research often finds a positive association between the prevalence of these boundaries and segregation levels across metropolitan areas, they do not explicitly focus on the physical barriers themselves and how their (dis)connectivity structures patterns of residential segregation.

A handful of studies have also examined specific types of boundaries and ethnoracial composition patterns in specific areas. In an analysis of Chicago, Noonan (2005) finds that differences between the share of blacks in adjacent areas is greater along physical barriers like parks, landmarks, railroads, state highways, major roads, and industrial corridors, rather than along census divisions that are not separated by such barriers, and Mitchell and Lee (2014) find an association, though weak, between socioeconomic differences and the presence of physical features, like rivers, parks, railroad tracks, and highways as neighborhood boundaries, rather than other divisions, such as streets, in Glasgow. Kramer (2017) finds that physical neighborhood boundaries, such as large roads, rivers, and railroad tracks, overlap with racial differentiation particularly in areas with histories of racial tension in Philadelphia.

Although these studies consider physical barriers, they do not incorporate connectivity, which is a key feature for understanding how physical barriers may affect segregation. A study by Le Goix (2005) that focuses exclusively on gated communities incorporates the notion of connectivity but does not find an effect of gated communities on racial segregation in Los Angeles. Le Goix (2005) explains that racial groups in the region generally separate along municipal boundaries instead and that many of the gated communities were built in unincorporated areas that eventually became their own municipalities or part of new ones. Grannis’ (1998, 2005) work gives most attention to physical barriers and road connectivity and demonstrates that racial variation in residential compositions tends to occur between rather than within t-communities in Chicago, Los Angeles, New York, and San Francisco. These studies, however, do not explicitly measure how these disconnected communities affect segregation levels.

To better understand how physical barriers facilitate the persistence of segregation, we build on the insights from this existing research to systematically examine how physical barriers influence residential segregation levels across several cities. Our analysis extends this research by examining multiple types of physical barriers that restrict connectivity in urban space; by considering the connectivity of these physical features, through bridges or pedestrian crossings; and by measuring how these features of the built environment change segregation levels. The goal of our paper is not to draw causal conclusions but rather to use improved measures of segregation to examine how physical barriers are related to segregation levels.

Below, we describe existing methods for measuring segregation and their shortcomings. Then, we describe the method that we use for measuring segregation, which was recently developed by Roberto (2015) and extends upon recent developments in measuring segregation. Specifically, the method allows us to consider physical barriers, integrates multiple groups and spatial proximity, and does not impose geographic units of analysis on the structure of segregation in a place. Next, we test our hypothesis that physical barriers facilitate segregation by comparing measures that incorporate the road connectivity between locations and those that do not across 20 U.S. cities. We examine patterns within cities and decompose our results by racial and ethnic group.

## RESIDENTIAL SEGREGATION MEASURES

While research implicates physical barriers in structuring residential segregation, the goal of our study is to directly examine how these features influence residential segregation levels. An analytic approach that compares residential segregation levels between measures that consider and do not consider such boundaries can shed light on this question. Residential segregation measures, however, have faced limitations in incorporating these features to assess this question.

Despite a long interest in segregation, few studies use measures that are consistent with the theoretical concept of segregation and its processes. Most studies focus on two of the five dimensions of segregation that Massey and Denton (1988) identify in their review of measures: evenness and exposure.[2] The dissimilarity index—which measures the extent to which two groups are evenly distributed across smaller spatial units (e.g., neighborhoods) within a larger spatial unit (e.g., a city)—remains the most widely used measure of evenness, and exposure measures the degree to which one group lives in similar spatial units as another group (Massey and Denton 1988).

Although these widely used measures are convenient to calculate with public data, they have notable shortcomings that limit our understanding of the mechanisms of segregation. First, traditional measures of segregation are limited to comparisons between two groups at a time, but, as the U.S. becomes increasingly multiethnic, segregation measures that incorporate multiple groups are increasingly important for understanding segregation patterns and trends (Reardon and Firebaugh 2002). Second, these measures are "aspatial"—they do not capture spatial concepts, such as where areas are located relative to each other, the relative size of segregated clusters, or the geographic extent of segregation patterns (White 1983; Morrill 1991; Reardon and O'Sullivan 2004; Brown and Chung 2006). Third, these measures do not consider variation in the geographic scale of spatial units and instead impose assumptions about which scales are relevant (Lee et al. 2008; Reardon et al. 2008; Fowler 2016). For example, a census tract with equal representation of each racial group as a city's overall population composition would reflect evenness, or low dissimilarity, but it could have distinct racial and ethnic enclaves within the tract, with residents living in racially homogeneous blocks. Thus, if blocks were the spatial unit of analysis, the area would appear highly segregated. In other words, the degree of segregation measured with traditional measures is sensitive to the size of geographic units, and these units are often defined by arbitrary administrative borders, like census-defined tracts or blocks (Grengs 2007; Spielman and Logan 2013).

[2] The other three dimensions are centralization, concentration, and clustering.

Scholars have noted these shortcomings for decades (Duncan and Duncan 1955; Openshaw and Taylor 1979; Massey and Denton 1988). To address these issues, researchers have developed many innovative solutions, including incorporating multiple ethnic groups into traditional measures and increasingly using entropy (i.e., information theory) indexes, which consider the diversity within a smaller spatial unit relative to the diversity of a larger aggregate spatial unit (for a review of these and other approaches, see Reardon and Firebaugh [2002]). Other recent developments, such as Reardon and O'Sullivan's (2004) multi-group spatial exposure index and spatial information theory index, incorporate spatial proximity and scale using the location of small spatial units and their proximities to one another (for reviews of other measures, see Wong [1993] and Reardon and O'Sullivan [2004]). Others incorporate multiple scales of segregation to examine how segregation changes depending on the spatial extent used to define a "local environment"—the area surrounding individuals' residential spaces. With this approach, a segregation measure is calculated for multiple distances from each residential space yielding a "segregation profile" (Lee et al. 2008; Reardon et al. 2008; Spielman and Logan 2013; Fowler 2015). By incorporating multiple ethnic groups, space, and scale into segregation measures, researchers have been able to compare patterns of segregation across groups and metropolitan areas (Lee et al. 2008; Reardon et al. 2008), identify neighborhoods (Spielman and Logan 2013), and examine which patterns are present at different geographic scales (Fowler 2016).

These developments have undoubtedly advanced our understanding of segregation, but these approaches do not integrate physical barriers. This is because measures that incorporate distance and spatial scale rely on *straight line distance*—the distance from Point A to Point B—without considering that spatial areas are often not connected in this way but rather through a road network. While Grannis' (1998) work on "t-communities"—streets connected by a tertiary intersection (i.e., streets connected by single lanes in each direction with no dividers)—emphasizes the effect of road connectivity on social interaction, its insights on the importance of road connectivity informs our approach. Physical barriers limit road connectivity, and we therefore extend upon this insight for incorporating physical barriers into a measure of segregation.

## RESEARCH DESIGN

We draw from publicly available population data from the 2010 decennial census (U.S. Census Bureau 2011) and the TIGER/Line shapefiles for blocks and roads (U.S. Census Bureau 2012).[3] The U.S. census subdivides the entire U.S. into several nested geographic units. We rely on information for census *blocks*—the smallest unit of census geography for which data are publicly available. Blocks are polygons that are typically bounded by street or road segments on each side and often correspond to the size of a city block in urban areas.[4] The Census groups blocks into *block groups*, which contain an average of 39 blocks and an average population of 1,500 people; block groups are grouped within *census tracts*—the most commonly used unit of analysis for constructing segregation measures and contain an average population of 4,000 individuals.

---

[3] In preparation for the 2010 census, Census Bureau employees walked virtually every street in the U.S. and used GPS technology to realign and update the geographic data for the road network, including adding new roads, renaming roads, or identifying roads to be deleted that no longer existed (U.S. Census Bureau 2012). The improvements to the TIGER/Line shapefiles make it unlikely that connected roads are disconnected in the data, or vice-versa (for more information about data accuracy, see U.S. Census Bureau [2012]).

For our analysis, we must first select which cities to explore more closely. Because the method we use is computationally intensive, we focus our analysis on 20 cities in the Rustbelt region – the former "industrial belt that extended from New England across New York, Pennsylvania, and West Virginia, through the Midwest to the banks of the Mississippi" (Sugrue 1996:6). Rustbelt cities have well-documented histories of residential segregation, reaching particularly high levels during the mid-twentieth century when these cities peaked in urban growth, and segregation has been the most persistent in these cities (Logan 2000). Concurrent with these legacies of segregation, policy and planning decisions that shaped the built environments of these cities, such as where to build interstate highways or public housing, were often made in response to racial tensions (Mohl 2002; Schindler 2015; Sugrue 1996). Although federal legislation has outlawed explicit forms of housing discrimination, many of these physical barriers remain intact and may or may not continue to facilitate segregation.

We focus our analysis on a set of 20 U.S. cities in the Rustbelt. We selected the 10 largest cities and 10 small-to-medium cities to capture a range of city sizes. For the largest city, New York, NY, we also analyze each of the five boroughs separately—Brooklyn, Bronx, Manhattan, Queens, and Staten Island. The remaining nine of the 10 largest Rustbelt cities are Chicago, IL; Philadelphia, PA; Indianapolis, IN; Columbus, OH; Detroit, MI; Baltimore, MD; Boston, MA; Washington, DC; and Milwaukee, WI. We also included the following six cities with populations between 250,000 and 500,000: Cleveland, OH; St. Louis, MO; Pittsburgh, PA; Cincinnati, OH; Newark, NJ; and Buffalo, NY. Lastly, we included the following four cities with populations between 100,000 and 250,000: Hartford, CT; New Haven, CT; Providence, RI; and Rochester, NY. Because the spatial organization of cities varies in other areas of the country, such as the West, we believe that there may be alternative, though perhaps related, spatial processes associated with the persistence of residential segregation in other cities that merit future analysis. While the findings that we present may not be generalizable to other cities, this selection of cities offers a starting point for future research.

### *Measuring Segregation*

The next step in our analysis is to implement a method for measuring segregation across these cities that incorporates multiple racial and ethnic groups, considers spatial proximity, measures segregation at multiple geographic scales, and allows us to consider how urban space is (dis)connected. To do this, we rely on a novel approach for measuring and analyzing segregation called Spatial Proximity and Connectivity (SPC) developed by Roberto (2015), which we describe in detail below.[5] The main advantage of this method is its flexibility to incorporate additional features of the spatial environment, which then allows us to compare across different measures that do or do not incorporate physical barriers.

[4] Blocks can represent spatial areas without residential land use, such as industrial areas, parks, or areas between railroad tracks and can have no population.
[5] We use R software and packages to implement this method (Csardi and Nepusz 2006; Kane et al. 2013; Bivand and Rundel 2014; Bivand et al. 2014; Neuwirth 2014; R Core Team 2014; Revolution Analytics and Weston 2014; Pebesma and Bivand 2015).

The first feature of the method incorporates multiple racial and ethnic groups, overcoming the limitation of many traditional segregation measures that only allow two-group comparisons. We use a segregation index developed by Roberto (2015) called the Divergence Index. The Divergence Index is based on relative entropy—an information theoretic measure also known as Kullback-Leibler (KL) divergence (Kullback 1987; Cover and Thomas 2006). It measures the difference between the composition of groups within a smaller spatial unit relative to the composition in a larger aggregate spatial unit. The Divergence Index differs from Reardon and O'Sullivan's (2004) Information Theory Index because it compares the population *compositions*, rather than the *diversity*, of smaller spatial units to the overall composition of a larger aggregate spatial unit (Roberto 2015). For our purposes, measuring the difference between compositions, rather than diversity per se, is favorable because it allows us to directly assess which groups are over- or under-represented relative to their overall proportions, whereas diversity measures lose information about specific groups and instead focus on variety or the relative quantity of groups. For example, in a city with a population composition comprised of 25 percent of group A and 75 percent of group B, an area of the city with the same population composition as the city—25 percent group A and 75 percent group B—would have the same diversity index as an area with the inverse population composition (75 percent of group A and 25 percent of group B)—a composition very different from that of the overall city and thus indicating segregation.

The Divergence Index can be interpreted as a measure of surprise—how surprising is the composition of smaller geographic areas, which we call "local environments," given the overall composition of a larger aggregate area of interest, which we call a "region." If all local environments within a region have the same composition as that of the region, the Divergence Index would equal 0 and would indicate no segregation, while higher values indicate more segregation. In this way, the Divergence Index is conceptually similar to the Dissimilarity Index and is consistent with the evenness dimension of segregation identified by Massey and Denton (1988). In the results that we present, we focus on segregation between whites, blacks, Hispanics, and Asians. The results including all major ethnoracial groups, however, are similar and are available upon request.

The second feature of the SPC method incorporates spatial proximity. Following developments from segregation measurement that incorporate geographic proximity (Reardon and O'Sullivan 2004; Fowler 2015), we use proximity-weights for population sizes within the local environments to weight the relative influence of nearby and more distant locations. We use a uniform proximity function in the results that we present, but the SPC method can accommodate other proximity weight functions.[6]

A third feature of the SPC method incorporates various geographic scales. Instead of using census tracts or other administratively-defined spatial units as the local environments, we use the intersections of roads and the distance, or "reach," from these intersections to define a set of overlapping local environments, similar to other recent approaches (Lee et al. 2008; Reardon et al. 2008; Speilman and Logan 2013). For a defined reach of the local environments (RLE), we calculate the Divergence Index for each intersection by comparing the composition of each intersection's local environment to the overall racial composition of the city population. More details on the construction of the Divergence Index are available in the Appendix.

[6] Others have used a two-dimensional biweight kernel proximity function (e.g., White 1983; Reardon and O'Sullivan 2004; Lee et al. 2008). Reardon and O'Sullivan (2004) do not recommend a specific function and conclude that results within cities would not be entirely sensitive to this choice.

We present results for RLEs ranging from .5 to 4 km (0.3 to 2.4 miles). The overall Divergence Index value presented for each city and each RLE distance is the population-weighted average of Divergence Index values for every intersection in the city for that RLE. The method also allows us to calculate the average degree of segregation experienced by residents in each ethnoracial group in each city. Thus, we can also examine if physical barriers influence segregation levels differently across groups. The population-weighted average of these group-specific results equals a city's overall Divergence Index. More details for this calculation are provided in the Appendix.

*Assessing Physical Barriers*

To assess how physical barriers influence residential segregation, we use the SPC method to construct each intersection's local environment using two different measures of distance—straight line distance and road distance. The straight line distance, which existing spatial measures of segregation use, measures the shortest distance between each pair of intersections in a city. Measuring segregation with the road distance requires linking the geographic and population data from the U.S. Census to estimate the racial composition of each intersection and constructing a road network using geographic information for all roads, including alleys and pedestrian walkways (e.g., walking paths through parks).

To link the geographic and population data, we use the TIGER/Line shapefiles for *faces* and *edges* to define the geographic boundaries of census blocks and the path of roads. *Faces* are polygons that represent areal features, such as blocks. Each face is assigned a permanent unique identifier by the Census Bureau. In most cases a block consists of a single face, but in some cases a block may contain two or more faces (e.g., if an alley subdivides a block). Each face is bounded by one or more edges. *Edges* are lines features, including road segments, and each edge has a unique identifier. Each edge is associated with two faces—one on each of its sides. The two end points of an edge are called *nodes*, and each has a unique identifier. A single node may be associated with multiple edges, such as a node that joins two road segments together.

For example, Figure 3 shows two blocks, the seven road segments that define their perimeters, and the intersections of the roads. Each of the blocks has one face, each road segment is an edge, and each end point of a road is a node. Roads that intersect have nodes in common. For example, node #65970117 in Figure 3 is an end point of edges #3701349, #3701194, and #3701194. This node is the shared intersection of Center Street and the two segments of Church Street.

[Figure 3 about here.]

We use the unique identifiers to identify the relationships between blocks, roads (including alleys and pedestrian walkways), and nodes and generate a list of the roads' edge identifiers and the node identifiers associated with each block. For each block, we identify the face identifier(s) associated with the block, find the edge identifiers for any roads that have the block's face identifier(s) listed as one of its sides, and collect all of the node identifiers for the end points of those roads. The result is a list of the edge identifiers and node identifiers associated with each block. We use this list for two purposes. First, we use the node identifiers associated with each block to assign a portion of each block's population to the nodes associated with it, thereby distributing the aggregate population of the block to these point locations. The procedure allows us to calculate the population count and composition in the local environment around each node. More details on this procedure are provided in the Appendix. Second, we use the list to construct the road network for each city. The road network is represented as a graph, with the nodes as the vertices and the road segments as the edges between them. Each edge of the graph is weighted by the length of the road segment it represents.

Next, we calculate the shortest distance between all pairs of nodes along the road network—our measure of the road distance.[7] Thus, the road distance measure captures the connectivity of roads and thereby reflects the excess distance created by physical barriers. Moreover, the physical barriers that the measure captures include gates, walls, cul-de-sacs, as well as highways, rivers, and railroad tracks without through-passes. If roads connected all intersections, then there would be no difference between the straight line and road network measures of distance. The presence of physical barriers will result in a difference between the relative size of a local environment constructed using road distance and a local environment constructed using straight line distance. The biggest differences will occur in areas where one or more nodes are not well connected to other nearby nodes, i.e., where physical barriers are present. Figure 4 illustrates the difference between a local environment constructed with the straight line distance and a local environment constructed with the road distance for one node. The node is located at the end of a dead-end street and is disconnected from the nearby nodes to the east, west, and south. This lack of road connectivity only affects the areas included in the node's local environment when it is constructed with the road distance. Figures 4a and 4b illustrate the distinct local environments depending on whether the straight line or road distance is used to construct them. Figure 4a shows the node's local environment constructed with the straight line distance using a reach of .5 km. All intersections within .5 km are included in the node's local environment. Figure 4b shows the node's local environment constructed with the road distance using the same reach of .5 km. The lack of connectivity to nearby nodes severely reduces the number of nodes that are included in the local environment.

If the node was connected to the nearby intersections to the east, west, or south, the two local environments would be much more comparable in scope. Figure 4c shows a counterfactual road network in which the focal node is connected to two proximate nodes, increasing the connectivity of roads from east to west. While there are 99 nodes included in the local environment constructed with the straight line distance (Figure 4a), there are only 12 in the original local environment constructed with the road distance (Figure 4b). When the two connections are added to the road network, the number of nodes included in the local environment increases from 12 to 58 (Figure 4c). Thus, connecting this one dead-end street to just two other nearby roads make the local environments much more comparable.

[7] We use the Dijkstra algorithm in the igraph R package (Csardi and Nepusz 2006) to calculate shortest paths.

[Figure 4 about here.]

In practice, it is not realistic to connect *all* intersections, and the impact of doing so also has comparatively less impact on a node's local environment than adding just one or two connections to nearby nodes. Appendix Figure A1 illustrates this trend. When we add connections one at a time from the focal node and every other node in the local environment using straight line distance to which it is not already directly connected via the road network, starting with the closest node (measured with the straight line distance), adding the first connection—between the focal node and the closest disconnected node—has the biggest impact on the size of its local environment, increasing the count of nodes from 12 to 40. The additional connections have smaller impact, especially after the second and third added connection.

The presence of physical barriers alone does not indicate that they have an impact on the level of segregation. For physical barriers to impact segregation levels, they must create greater separation between areas with different ethnoracial groups. To analyze this, we measure segregation using local environments constructed with each type of distance measure such that the measure of segregation is a function of these distance measures. We calculate two distinct segregation measures for each intersection's local environment at each specified RLE: the *straight line distance Divergence Index* and the *road distance Divergence Index*. We then compare the two sets of segregation results. For example, for a RLE of 1 km, at each intersection in a city, we calculate the Divergence Index for the composition of the spatial area within a 1 km radius of the intersection and compare this composition to the composition of the region; this is the straight line distance Divergence Index. We also calculate the Divergence Index for the composition of the spatial area within 1 km via the road network from the intersection and compare this composition to the composition of the region; this is the road distance Divergence Index.

When there is no difference between the road distance and straight line distance Divergence Indexes for an intersection's local environment but physical barriers are present, this indicates that physical barriers do not structure the spatial pattern of segregation in this local environment. This can occur where the racial compositions of the local environments measured by each type of distance measure are identical. A road distance Divergence Index lower than the straight line distance Divergence Index for a local environment occurs when a physical barrier separates a spatial area that has a composition similar to the region (i.e., lower segregation) from an area with a composition that differs from the region (i.e., higher segregation). If the road distance Divergence Index for a local environment is higher than the straight line distance Divergence Index, this indicates that physical barriers play a role in increasing segregation for residents of that intersection.

To evaluate the extent to which physical barriers structure the overall segregation of each city, we compare the overall Divergence Indexes for each city and for each type of distance measure at varying RLEs using the population-weighted averages of the local Divergence Indexes for all intersections in each city. The magnitude of the difference between the segregation measures indicates how much of an impact physical barriers have on segregation.

Even small positive differences between the straight line and road distance Divergence Indexes indicate that physical barriers facilitate greater separation between ethnoracial groups and higher levels of segregation in a city. This is because the overall city-level segregation results for each RLE are the population-weighted average of segregation in the local environments of each intersection, including areas where there are no physical barriers. Given that the overall difference reflects the average difference for all intersections within the city, any overall difference—even if small in value—is meaningful. In many areas within cities, the local environments constructed with road network and straight line distance will have similar compositions to each other and will thus yield zero differences in segregation. Because the magnitude of the differences can appear small, we also examine areas of cities where distinct ethnoracial groups are proximate to each other to assess the role of physical barriers in these areas on the Divergence Indexes.

For locations near the municipal boundary of the city, we limit the local environments to be within the municipal boundary of the city by only including locations within the specified RLE (either by road network or straight line distance) that are in the city. For example, if the RLE is 1 km, the local environment of an intersection within 1 km of the municipal boundary does not include residents across the municipal boundary even though they are within 1 km of the intersection.[8]

## RESULTS

We first describe the cities that we examine in our study. Table 1 displays the population size and share of non-Hispanic whites, non-Hispanic blacks, Hispanics, and Asians for the 20 cities based on the 2010 decennial census. The table illustrates how most of the cities that we include in our analysis are not majority white. Boston, Columbus, Indianapolis, Pittsburgh, and two boroughs of New York City—Manhattan and Staten Island—have much larger shares of non-Hispanic whites compared to other race groups. In contrast, Baltimore, Cleveland, Detroit, Newark, and Washington have much larger shares of blacks compared to other race groups in their respective cities. In Hartford and the Bronx, Hispanics comprise the largest share of the populations. Asians are a minority share of the population in all of the cities that we include in our analysis.

[Table 1 about here.]

Figure 5 displays four multi-group measures of segregation for Asian, black, Hispanic, and white residents for each city: the Index of Dissimilarity (Reardon and O'Sullivan 2004), the Normalized Exposure Index (Reardon and Firebaugh 2002), and the aspatial and spatial versions of the Divergence Index (Roberto 2015). The first two measures are extensions of traditionally-used measures of segregation, and the first three measures use census tracts as the units of analysis. The fourth measure—the spatial Divergence Index—is the population-weighted average of the road network Divergence Indexes for each intersection in a city based on a .5 km RLE.

[8] Local environments can span bodies of water, such as a rivers and lakes, and will include the population on the other side of the water if it is within the RLE (via the road network or straight line distance). Where the city is bordered by another country (e.g., where Detroit borders Canada), the local environments for all intersections do not cross the U.S. border.

[Figure 5 about here.]

The Dissimilarity and Normalized Exposure Indexes both indicate that Chicago, Detroit, Baltimore, Milwaukee, Philadelphia, Cleveland, Washington, St. Louis, and Brooklyn are among the most segregated cities of the group, while Indianapolis, Columbus, Harford, New Haven, Rochester, and Providence, as well as Staten Island and the Bronx have relatively low levels of segregation on both of these measures. A high Dissimilarity Index indicates that a large share of residents would need to move to a different tract to generate an even distribution of ethnoracial groups across tracts, and a high Normalized Exposure Index indicates that the ethnoracial groups have little exposure to each other in the tracts where they currently live. The aspatial Divergence Index reveals similar patterns to the two measures except Detroit and Baltimore have relatively less segregation and Boston has relatively more segregation by this measure. The spatial Divergence index reveals similar patterns as the aspatial version of the measure, except Indianapolis has relatively higher segregation by this measure.

These differences across measures stem from the aspatial ways in which the first three indexes are measured and the distinct features that these measures capture. For example, most tracts in Detroit are primarily black in a city with a population that is 82 percent black. The relatively small proportion of Hispanics tend to reside in separate tracts in and around "Mexicantown" in southwest Detroit while the relatively small proportion of whites tend to reside in tracts downtown, leading to relatively high levels of tract segregation by the Dissimilarity and Normalized Exposure Indexes.[9] But because the areas with large white or Hispanic populations are surrounded by areas with a different population composition, most local environments with a .5 km RLE are similar to the overall city composition and therefore have a low spatial Divergence Index.

*Physical Barriers and Segregation*

Next, we compare segregation measures incorporating road distance to those that use straight line distance for each of the cities to examine how physical barriers influence segregation levels. We assess overall segregation levels for each city for five RLEs: .5 km, 1 km, 2 km, 3 km, and 4 km, which each approximate the area of a single census tract, three to four census tracts, larger aggregate units often used as neighborhood statistical areas, and even larger aggregate units that conform to large regions of cities, respectively. Figure 6 presents the percent differences between the road distance Divergence Index relative to the straight line distance Divergence Index for each city. The segregation levels are presented in Appendix Table A1 and the differences and percent difference are presented in Appendix Tables A2 and A3.

[Figure 6 about here.]

The results indicate that the differences between the two measures of segregation vary depending on the RLE and across cities. The meaning of each RLE, of course, varies across cities depending on the geographic size and population density of the city itself. At an RLE of .5 km, the average difference across all cities is about 4 percent (see Table A2). Cincinnati, Columbus, Hartford, Indianapolis, and Pittsburgh have the largest percent differences at this RLE—all greater than a 6 percent increase. This represents an increase of .02 to .03 in the overall level of segregation levels in these cities (see Table A3 and Figure A2).

[9] The Dissimilarity Index is upwardly biased when the population includes small groups (e.g., Winship 1977; Carrington and Troske 1997).

Many of the differences are relatively small at this scale, indicating that the racial compositions of most local environments constructed with road network distance are similar to those constructed with straight line distance at this scale. This is not surprising given the average measures include all locations in a city but most locations in a city do not include physical barriers in their local environments where there are no physical barriers. Many of the spatial features that create large differences at smaller RLEs, such as dead-end streets, tend to affect connectivity within small local areas rather than for entire sections of a city and may not have a great impact on the overall difference but still indicate that physical barriers affect segregation. To observe *any* overall difference, even if seemingly small in value, indicates that there is greater separation between ethnoracial groups in the city resulting from physical barriers.

*Variation by Reach of Local Environments*

Greater differences are evident in all cities at greater reaches. Using an RLE of 2 km, the differences between the segregation measures increase across all cities with Baltimore and Pittsburgh having the highest percent increases in segregation levels of 35 percent each. The spatial pattern of ethnoracial clustering in these two cities resembles a patchwork design, with racial or ethnic enclaves throughout the city, in contrast to other cities where the ethnoracial clustering encompasses large sections of the city. The large differences between the segregation measures in these cities suggests that physical barriers tend to mark the borders of segregated clusters in these cities. In Cleveland, Detroit, St. Louis, and Washington, on the other hand, the differences between the Divergence Indexes remains relatively small, and, in these cities, racial groups are separated in large residential clusters, such as the pattern of ethnoracial clustering dividing the north and south sides of St. Louis or Cleveland. Thus, for most intersections in these cities, the local environments measured by straight line and road distances have similar compositions. Where these ethnoracial clusters converge is where differences are more likely. While physical barriers may indeed mark the clusters' borders, the relatively fewer instances of where these groups meet explains why physical barriers influence the Divergence Index less on average within these cities.

While percent differences between the segregation measures increases with the RLE, segregation levels tend to decrease with the RLE. In many cities, the maximal difference between the two segregation measures occurs in local environments that are not necessarily the largest. For example, in Hartford, Manhattan, and New Haven, the difference is highest in local environments with a 1 km reach, which suggests that physical barriers have a greater average effect on structuring segregation within 1 km of intersections rather than at greater distances. In other cities—Baltimore, Boston, Cincinnati, Columbus, Staten Island, Pittsburgh, and Rochester—the difference is highest in local environments with a 2 km reach. This likely reflects the sizes of the ethnoracial clusters in the cities. If physical barriers mark the boundaries of ethnoracial clusters, then the difference between segregation measures would be greatest when the reach using straight line distance is about the same as the radius of a typical cluster. At that reach, local environments will include the physical barrier when measured with straight line distance, but would exclude it when measured with road distance. When the reach using straight line distance is much larger than the radius of a typical cluster, the difference will gradually diminish.

Altogether, these results provide evidence that physical barriers increase the degree of segregation across cities on average. Nonetheless, the degree to which they matter varies by the size of local environments for which we measure segregation. Further, the differences generally increase when we consider larger RLEs.

*Local Effects of Physical Barriers*

Figure 7 illustrates how physical barriers can structure local patterns of residential segregation using the example of Baltimore, which is 63 percent black, 4 percent Hispanic, 28 percent white, and 2 percent Asian. The remaining 3 percent are another ethnoracial group. In Figure 7, we highlight an area of Baltimore—the neighborhoods of Waverly and Guilford, which are divided by Greenmount Avenue. Waverly is a poor, predominantly black neighborhood on the east side of Greenmount Avenue, and Guilford is a wealthy, predominantly white neighborhood on the west side of Greenmount Avenue (Schindler 2015). Armborst et al. (2015) describe Greenmount Avenue between 33rd Street and Cold Spring Lane as a wall used to disconnect one side from the other. For example, among the streets that intersect Greenmount Avenue between 33rd Street and Cold Spring Lane, very few streets cross Greenmount Avenue to connect the two neighborhoods.

The map illustrates this lack of road access between Guilford and Waverly crossing Greenmount Avenue. Whereas Waverly contains a dense network of roads within it, the roads connecting it with Guilford are far less dense. As a result, we find differences between the road distance and straight line distance Divergence Indexes. The intersections with darker shaded dots indicate relatively larger differences between the road distance and straight line distance Divergence Indexes for each intersection's local environment using an RLE of .5 km, and the lightest shaded dots indicate negative differences. The prevalence of darker points in the map indicates how the lack of road connectivity between Guilford and Waverly exacerbates segregation for individuals living within both of these neighborhoods.

When we compare the average segregation levels for each neighborhood, we find that the road distance Divergence Index is 12 percent and 34 percent greater than the straight line distance Divergence Index in Guilford and Waverly, respectively. Segregation levels increase from 0.64 to 0.73 in Guilford and from 0.03 to 0.04 in Waverly when measured with the road distance Divergence Index rather than the straight line distance Divergence Index. These differences suggest that the residents of the wealthy, predominantly white neighborhood of Guilford are even more segregated from their nearby neighbors than standard estimates would detect when we take the physical barriers into account. Nonetheless, our estimate of the extent to which the lack of connectivity across Greenmount Avenue exacerbates segregation may be conservative since our method incorporates the connectivity provided by walking paths across Greenmount Avenue between Waverly and Guilford, despite the presence of bollards and one-way streets that inhibit vehicle traffic across these neighborhoods (Mikin 2012).

[Figure 7 about here.]

*Differences by Race*

The differences discussed thus far reflect the averages for all residents in each city. When we examine the differences separately for white, black, Hispanic, and Asian residents, we find striking variation in race-specific differences between the road distance and straight line distance Divergence Indexes. The results demonstrate that, even in cities that have relatively small differences on average for all groups, the differences are often much larger for particular ethnoracial groups. Table 2 presents the population-weighted averages across cities of the differences between the road distance and straight line Divergence Indexes by ethnoracial group. The results indicate that at smaller RLEs, the differences are much larger for Asians relative to other groups, but at larger RLEs, the differences are greater for blacks relative to other groups. In other words, when local environments are defined as small geographic areas, physical barriers result in higher levels of segregation experienced by Asians, but for local environments that span larger areas, physical barriers appear to impact blacks the most—increasing the degree of segregation that they experience. These differences may stem from distinct types of physical barriers that structure the segregation patterns of these groups, as well as differences in the geographic scale of racial and ethnic clustering of these groups, with blacks tending to live in areas with more of a macro-scale pattern of clustering and Asians tending to live in areas with a more micro-scale pattern.

[Table 2 about here.]

Nonetheless, differences across ethnoracial groups vary greatly across cities and by the RLE. Figure 8 shows the differences between the segregation measures for each ethnoracial group in each city, and Figure A3 illustrates the percent differences between the measures.

For example, in Baltimore and Newark there are larger positive differences between the segregation measures for white residents compared to the overall average for all residents in these cities. In Boston, Buffalo, Indianapolis, and Pittsburgh, there are larger positive differences between the segregation measures for black residents compared to the overall average for all residents in these cities. In Hartford, the race-specific differences depend on the RLE: there are larger positive differences for white residents in smaller local environments and for black residents in larger local environments. In Cleveland, Detroit, and Rochester, Asian residents are a relatively small proportion of the city population, but, nonetheless, there are much larger positive differences between the segregation measures in their local environments with reaches of .5 or 1 km compared to the overall average for all residents at these RLEs. These results suggest that physical barriers increase segregation levels for some ethnoracial groups in some cities but are associated with smaller changes in segregation for others.

[Figure 8 about here.]

## DISCUSSION AND CONCLUSIONS

While earlier work implicated physical barriers in facilitating residential segregation, our study draws on recent advances in the measurement of segregation to offer a direct examination of how these spatial boundaries relate to residential segregation levels. With this research, we seek to bridge the theoretical concept of segregation with its operationalization—following what Duncan and Duncan (1955) advocated for decades ago—by building on recent developments in segregation measures to incorporate multiple racial and ethnic groups, spatial proximity, and geographic scale and utilizing a novel method that allows us to incorporate physical barriers. We compare segregation levels using measures that incorporate these features with measures that imply the absence of these physical barriers (i.e., straight line distance measures), and we find that, in general, physical barriers increase levels of segregation across these cities. The findings also highlight important sources of variation in understanding the effect of physical barriers—ethnoracial groups, RLEs considered, and within and across cities. Altogether, the results imply that without these barriers in place, segregation would be lower and residential integration more likely.

Although legislation passed nearly 50 years ago ended explicit racial discrimination in real estate and lending practices and race relations have certainly improved since then, residential segregation levels remain persistently high across the U.S. Our study contributes to our understanding of segregation by demonstrating how physical barriers separate urban spaces in ways that increase residential segregation. These barriers often provide clear divisions between spaces that yield agreement among residents, real estate agents, and other institutional actors and influence residential sorting patterns (Ananat 2011; Bader and Krysan 2015). Consequently, the presence of these barriers can result in distinct social conditions and experiences for individuals on different sides of them (Besbris, et al. 2015). Weaker boundaries, like streets or landmarks, offer the *possibility* for integration whereby residents or real estate agents can reconstruct neighborhood boundaries, identities, and reputations, but physical barriers limit such integration.

Altogether, our results provide direct evidence that physical barriers increase residential segregation levels across several U.S. cities. Existing explanations of the persistence of residential segregation focus less on how space itself can contribute to the perpetuation of segregation. Considering the spatial dimensions of segregation, including features of the built environment, advances theoretical accounts of segregation. These spatial features shape patterns of residential sorting and influence the mechanisms that perpetuate segregation—such as socioeconomic differences, residential preferences, and housing market discrimination.

While we believe that our results provide convincing evidence that physical barriers reinforce or exacerbate segregation and limit integration, our ability to draw causal conclusions is limited. We do not examine changes in segregation levels over time with the introduction or removal of physical barriers. Our results suggest that these barriers increase segregation levels compared to if they were absent, but it is possible that the removal of a barrier may lead to alternative patterns of residential sorting that nonetheless maintain residential segregation. An analysis over time with historical data that examines both the creation and removal of such barriers would shed light on individual and institutional processes that generate racial segregation. Nonetheless, our results show that the continued existence of these boundaries continues to reinforce segregation today.

Our analysis examines a limited set of cities with relatively similar histories of urbanization, long histories of residential segregation, and relatively large black populations, but our findings and approach raise new questions that can enhance our understanding of mechanisms contributing to residential segregation. How do physical barriers structure residential segregation in cities with distinct histories of urban growth and residential segregation? For example, physical barriers may have strong roles in cities with seemingly historically low levels of segregation. We also find important variation across and within the cities that we examined (see Appendix Tables A1 and A2). In some cities, physical barriers may be more prevalent and associated with segregation in one section of a city, but there may be fewer or they may matter less in another section of the city. Such variation has important implications for understanding the consequences of segregation and their disparate impact on residents. Future research can integrate these results with additional data sources to further investigate the historical, political, and social factors responsible for the variation.

Methodologically, we use an innovative approach that advances the study of segregation by incorporating a variety of dimensions that previous measures do not consider (Roberto 2015). These dimensions allow us to assess how physical barriers are related to segregation, shedding new light on how we understand segregation and how it varies across space. While this approach significantly improves upon even the most advanced measures of segregation, which use straight line distance to measure the proximity of spaces in a city, there are even more avenues for incorporating spatial features into measures of segregation. First, we consider bridges and underpasses as connected roads, but these types of areas can also separate space if they are desolate or poorly lit, for example. In this case, they can perpetuate residential segregation rather than bridging spaces together. Thus, developing a systematic measure of the quality of connective roads and barriers could further enhance our understanding of how physical barriers structure residential segregation, and new methods for systematic social observation using Google Street View could provide a starting point for developing such a measure (Hwang and Sampson 2014). Second, we focus on where residents live, but further analysis could examine how physical barriers affect segregation in individuals' activity spaces with data on the locations where individuals spend time and interact with others.

Despite the further questions that this study poses, the findings have implications for reducing segregation. Urban design and planning that connects racially segregated areas separated by physical barriers can promote racial integration. For example, a recent initiative called Reimagining the Civic Commons is funding demonstration projects in five cities to revitalize and connect public spaces, including converting an abandoned elevated railroad into a park with walking paths (http://www.civiccommons.us/). Such an effort would connect the areas on both sides of the railroad tracks, having the potential to promote residential integration. Removing such barriers or the inequalities on either side of them may not guarantee residential integration, but it can weaken the barriers, make them negotiable, and therefore create the possibility for integration, which otherwise would not exist.

**TABLES**

**Table 1.** Population Compositions for Cities

| | Population Count (thousands) | Population Composition | | | | Land Area (square kilometers) |
|---|---|---|---|---|---|---|
| | | Non-Hispanic White | Non-Hispanic Black | Hispanic | Asian | |
| Baltimore | 621.0 | .28 | .63 | .04 | .02 | 209.6 |
| Boston | 617.6 | .47 | .22 | .17 | .09 | 125.0 |
| Buffalo | 261.3 | .46 | .37 | .11 | .03 | 104.6 |
| Chicago | 2,695.6 | .32 | .32 | .29 | .05 | 589.6 |
| Cincinnati | 296.9 | .48 | .45 | .03 | .02 | 201.9 |
| Cleveland | 396.8 | .33 | .52 | .10 | .02 | 201.2 |
| Columbus | 787.0 | .59 | .28 | .06 | .04 | 562.5 |
| Detroit | 713.8 | .08 | .82 | .07 | .01 | 359.4 |
| Hartford | 124.8 | .16 | .35 | .43 | .03 | 45.0 |
| Indianapolis | 820.4 | .59 | .27 | .09 | .02 | 936.1 |
| Milwaukee | 594.8 | .37 | .39 | .17 | .03 | 249.0 |
| New Haven | 129.8 | .32 | .33 | .27 | .05 | 48.4 |
| New York | 8,175.1 | .33 | .23 | .29 | .13 | 783.8 |
| Brooklyn | 2,504.7 | .36 | .32 | .20 | .10 | 183.4 |
| Bronx | 1,385.1 | .11 | .30 | .54 | .03 | 109.0 |
| Manhattan | 1,585.9 | .48 | .13 | .25 | .11 | 59.1 |
| Queens | 2,230.7 | .28 | .18 | .28 | .23 | 281.1 |
| Staten Island | 468.7 | .64 | .09 | .17 | .07 | 151.2 |
| Newark | 277.1 | .12 | .50 | .34 | .02 | 62.6 |
| Philadelphia | 1,526.0 | .37 | .42 | .12 | .06 | 347.3 |
| Pittsburgh | 305.7 | .65 | .26 | .02 | .04 | 143.4 |
| Providence | 178.0 | .38 | .13 | .38 | .06 | 47.7 |
| Rochester | 210.6 | .38 | .40 | .16 | .03 | 92.7 |
| St. Louis | 319.3 | .42 | .49 | .03 | .03 | 160.3 |
| Washington | 601.7 | .35 | .50 | .09 | .03 | 158.1 |

**Table 2.** Difference between Road Distance and Straight Line Distance Divergence Indexes by Ethnoracial Group (population weighted mean across all cities)

| | Reach of Local Environments | | | | |
|---|---|---|---|---|---|
| | .5 km | 1 km | 2 km | 3 km | 4 km |
| White | .018 | .026 | .032 | .034 | .033 |
| Black | .016 | .027 | .043 | .047 | .049 |
| Hispanic | .018 | .027 | .040 | .042 | .041 |
| Asian | .024 | .035 | .035 | .036 | .033 |
| Mean | .018 | .027 | .037 | .040 | .040 |

**Table A1.** Road Distance and Straight Line Distance Divergence Indexes by City

| | Straight Line Distance Divergence Index | | | | | | | | | | Road Distance Divergence Index | | | | | | | | | |
|---|---|---|---|---|---|---|---|---|---|---|---|---|---|---|---|---|---|---|---|---|
| | Reach of Local Environments | | | | | | | | | | Reach of Local Environments | | | | | | | | | |
| | .5 km | | 1 km | | 2 km | | 3 km | | 4 km | | .5 km | | 1 km | | 2 km | | 3 km | | 4 km | |
| | Mean | SD | Mean | SD | Mean | SD | Mean | SD | Mean | SD | Mean | SD | Mean | SD | Mean | SD | Mean | SD | Mean | SD |
| Baltimore | .46 | .38 | .37 | .34 | .21 | .23 | .11 | .15 | .06 | .10 | .48 | .38 | .41 | .36 | .29 | .28 | .17 | .20 | .10 | .15 |
| Boston | .48 | .37 | .40 | .34 | .28 | .23 | .20 | .15 | .15 | .13 | .50 | .38 | .44 | .36 | .34 | .29 | .25 | .19 | .18 | .15 |
| Buffalo | .50 | .29 | .43 | .27 | .32 | .24 | .23 | .21 | .16 | .17 | .52 | .30 | .46 | .27 | .36 | .25 | .28 | .23 | .21 | .20 |
| Chicago | .87 | .43 | .81 | .41 | .70 | .37 | .60 | .34 | .52 | .32 | .88 | .43 | .84 | .42 | .75 | .38 | .66 | .36 | .58 | .34 |
| Cincinnati | .35 | .28 | .28 | .25 | .18 | .21 | .12 | .17 | .08 | .14 | .38 | .29 | .33 | .27 | .23 | .23 | .16 | .20 | .11 | .17 |
| Cleveland | .56 | .31 | .52 | .27 | .45 | .23 | .41 | .19 | .37 | .16 | .57 | .31 | .54 | .28 | .47 | .25 | .43 | .22 | .40 | .18 |
| Columbus | .33 | .30 | .28 | .26 | .22 | .19 | .19 | .16 | .16 | .13 | .35 | .32 | .31 | .29 | .26 | .23 | .22 | .18 | .18 | .15 |
| Detroit | .37 | .68 | .34 | .63 | .29 | .54 | .24 | .47 | .21 | .38 | .37 | .68 | .35 | .65 | .31 | .58 | .27 | .51 | .23 | .45 |
| Hartford | .34 | .31 | .25 | .23 | .17 | .15 | .12 | .12 | .06 | .07 | .36 | .33 | .30 | .29 | .22 | .22 | .15 | .16 | .10 | .11 |
| Indianapolis | .38 | .34 | .33 | .30 | .27 | .25 | .23 | .20 | .20 | .15 | .41 | .36 | .36 | .33 | .31 | .28 | .27 | .24 | .23 | .20 |
| Milwaukee | .68 | .32 | .63 | .32 | .54 | .30 | .45 | .26 | .36 | .21 | .69 | .32 | .65 | .32 | .59 | .32 | .51 | .28 | .42 | .24 |
| New Haven | .38 | .25 | .27 | .19 | .12 | .12 | .07 | .09 | .04 | .06 | .41 | .26 | .32 | .22 | .15 | .13 | .09 | .10 | .06 | .08 |
| New York | | | | | | | | | | | | | | | | | | | | |
| Brooklyn | .68 | .31 | .59 | .28 | .46 | .25 | .37 | .23 | .29 | .21 | .70 | .32 | .62 | .29 | .50 | .25 | .41 | .24 | .33 | .23 |
| Bronx | .30 | .37 | .25 | .30 | .18 | .21 | .12 | .16 | .07 | .08 | .31 | .39 | .27 | .33 | .20 | .23 | .15 | .19 | .10 | .14 |
| Manhattan | .49 | .38 | .41 | .30 | .33 | .23 | .25 | .20 | .20 | .18 | .51 | .41 | .44 | .33 | .35 | .25 | .28 | .21 | .23 | .18 |
| Queens | .62 | .47 | .56 | .46 | .45 | .43 | .35 | .39 | .29 | .35 | .63 | .47 | .58 | .47 | .48 | .44 | .40 | .41 | .32 | .37 |
| Staten Island | .29 | .31 | .25 | .23 | .18 | .12 | .14 | .09 | .12 | .07 | .31 | .34 | .27 | .27 | .21 | .17 | .16 | .11 | .14 | .09 |
| Newark | .49 | .31 | .44 | .28 | .33 | .20 | .23 | .16 | .13 | .13 | .51 | .32 | .46 | .29 | .37 | .24 | .27 | .18 | .18 | .15 |
| Philadelphia | .63 | .36 | .55 | .33 | .43 | .25 | .35 | .20 | .28 | .18 | .65 | .37 | .58 | .34 | .46 | .28 | .38 | .22 | .32 | .20 |
| Pittsburgh | .33 | .38 | .25 | .31 | .15 | .20 | .09 | .11 | .05 | .06 | .37 | .40 | .30 | .35 | .20 | .26 | .13 | .19 | .08 | .10 |
| Providence | .30 | .23 | .24 | .21 | .16 | .15 | .09 | .09 | .04 | .05 | .32 | .24 | .26 | .22 | .19 | .18 | .12 | .12 | .07 | .08 |
| Rochester | .34 | .26 | .28 | .21 | .19 | .15 | .11 | .11 | .06 | .09 | .36 | .27 | .32 | .25 | .24 | .18 | .16 | .14 | .10 | .11 |
| St. Louis | .49 | .33 | .43 | .33 | .36 | .32 | .30 | .28 | .25 | .24 | .51 | .33 | .46 | .33 | .38 | .33 | .33 | .30 | .28 | .27 |
| Washington | .54 | .30 | .49 | .30 | .41 | .30 | .34 | .29 | .26 | .25 | .55 | .30 | .51 | .30 | .44 | .31 | .38 | .30 | .31 | .28 |

**Table A2.** Percent Difference between Road Distance and Straight Line Distance Divergence Indexes by City

| | Reach of Local Environments | | | | |
|---|---|---|---|---|---|
| | .5 km | 1 km | 2 km | 3 km | 4 km |
| Baltimore | 5.0 | 10.8 | 34.6 | 55.5 | 59.1 |
| Boston | 5.3 | 8.6 | 19.0 | 23.7 | 24.5 |
| Buffalo | 3.6 | 7.1 | 14.8 | 22.3 | 30.8 |
| Chicago | 1.7 | 3.1 | 6.6 | 9.5 | 12.4 |
| Cincinnati | 7.1 | 16.6 | 29.0 | 37.7 | 39.5 |
| Cleveland | 2.1 | 3.9 | 5.4 | 5.8 | 6.9 |
| Columbus | 6.9 | 10.0 | 14.5 | 16.2 | 16.7 |
| Detroit | 1.9 | 2.9 | 7.5 | 9.6 | 12.6 |
| Hartford | 7.9 | 18.5 | 24.4 | 27.9 | 57.8 |
| Indianapolis | 7.1 | 10.3 | 11.7 | 14.4 | 17.5 |
| Milwaukee | 1.8 | 3.2 | 7.9 | 13.5 | 16.5 |
| New Haven | 6.3 | 18.1 | 29.6 | 22.1 | 45.5 |
| New York | | | | | |
| Brooklyn | 2.5 | 4.8 | 8.6 | 10.5 | 14.7 |
| Bronx | 4.5 | 8.4 | 14.9 | 28.3 | 49.9 |
| Manhattan | 4.1 | 7.1 | 8.0 | 11.2 | 12.7 |
| Queens | 2.7 | 4.3 | 8.1 | 12.0 | 13.0 |
| Staten Island | 5.4 | 9.3 | 16.1 | 16.3 | 17.0 |
| Newark | 2.8 | 5.2 | 11.2 | 20.2 | 36.4 |
| Philadelphia | 3.2 | 5.4 | 8.7 | 10.1 | 15.8 |
| Pittsburgh | 10.5 | 20.1 | 34.8 | 46.6 | 60.1 |
| Providence | 5.8 | 9.4 | 19.9 | 43.1 | 80.7 |
| Rochester | 6.4 | 13.7 | 28.5 | 46.3 | 57.3 |
| St. Louis | 3.7 | 6.0 | 5.8 | 9.7 | 14.1 |
| Washington | 2.0 | 3.1 | 6.3 | 11.3 | 18.6 |
| Mean | 4.4 | 8.4 | 15.0 | 20.9 | 29.1 |
| SD | 2.4 | 5.5 | 9.9 | 14.6 | 21.2 |

**Table A3.** Difference between Road Distance and Straight Line Distance Divergence Indexes by City

| | Reach of Local Environments | | | | |
|---|---|---|---|---|---|
| | .5 km | 1 km | 2 km | 3 km | 4 km |
| Baltimore | .023 | .040 | .074 | .061 | .037 |
| Boston | .025 | .035 | .054 | .047 | .036 |
| Buffalo | .018 | .031 | .047 | .051 | .050 |
| Chicago | .014 | .025 | .046 | .057 | .064 |
| Cincinnati | .025 | .046 | .052 | .044 | .032 |
| Cleveland | .012 | .020 | .024 | .024 | .025 |
| Columbus | .023 | .028 | .033 | .030 | .026 |
| Detroit | .007 | .010 | .022 | .023 | .026 |
| Hartford | .027 | .047 | .042 | .033 | .035 |
| Indianapolis | .027 | .034 | .032 | .033 | .035 |
| Milwaukee | .012 | .020 | .043 | .060 | .060 |
| New Haven | .024 | .049 | .034 | .016 | .018 |
| New York | | | | | |
| Brooklyn | .017 | .028 | .040 | .039 | .043 |
| Bronx | .014 | .021 | .026 | .033 | .034 |
| Manhattan | .020 | .029 | .026 | .028 | .026 |
| Queens | .016 | .024 | .036 | .043 | .037 |
| Staten Island | .016 | .023 | .029 | .023 | .020 |
| Newark | .014 | .023 | .037 | .046 | .047 |
| Philadelphia | .020 | .030 | .037 | .035 | .044 |
| Pittsburgh | .035 | .051 | .051 | .042 | .031 |
| Providence | .017 | .023 | .032 | .037 | .030 |
| Rochester | .022 | .039 | .054 | .051 | .036 |
| St. Louis | .018 | .026 | .021 | .029 | .035 |
| Washington | .011 | .015 | .026 | .038 | .049 |
| Mean | .018 | .029 | .037 | .037 | .035 |
| SD | .007 | .012 | .014 | .015 | .014 |

## FIGURES

**Figure 1.** Stylized Example of How Cul-de-Sacs Serve as Physical Barriers

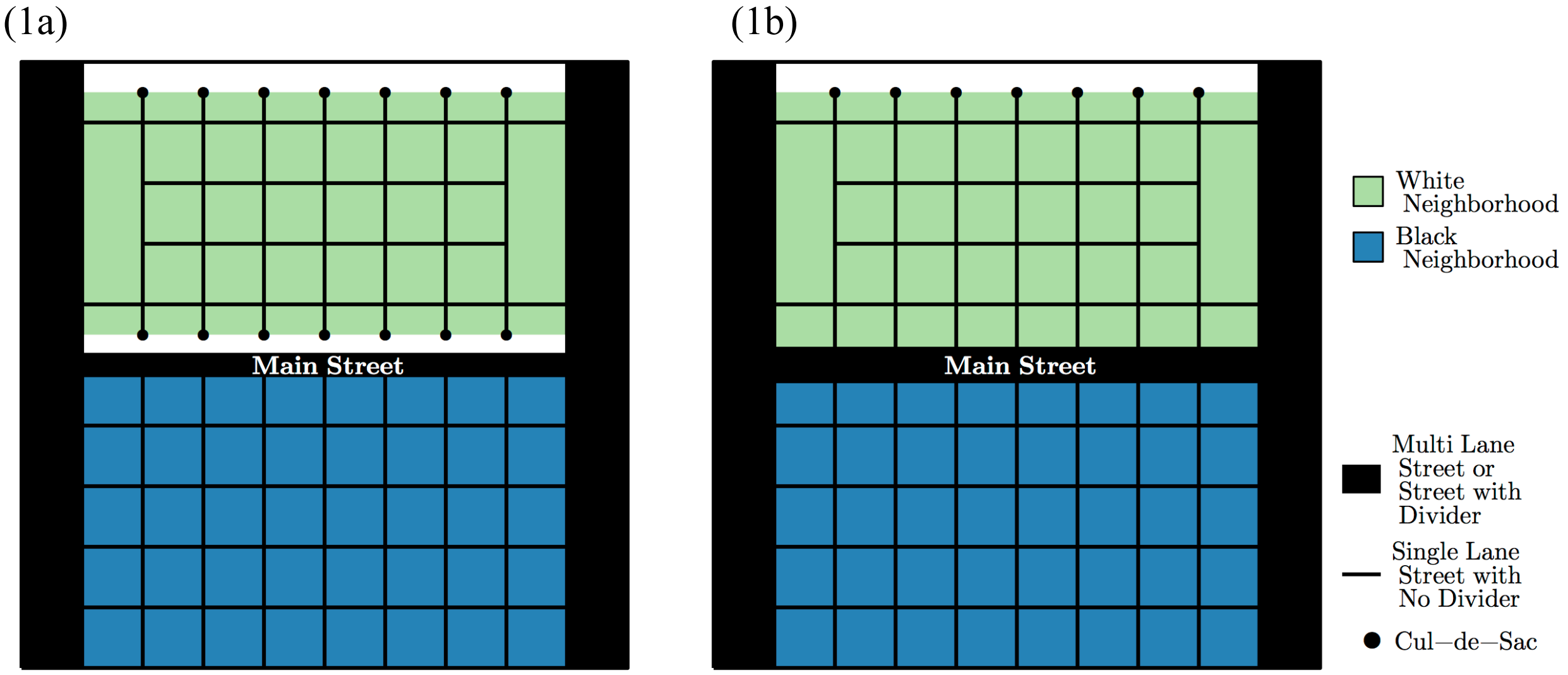

**Figure 2.** Stylized Example of How Railroad Tracks Serve as Physical Barriers

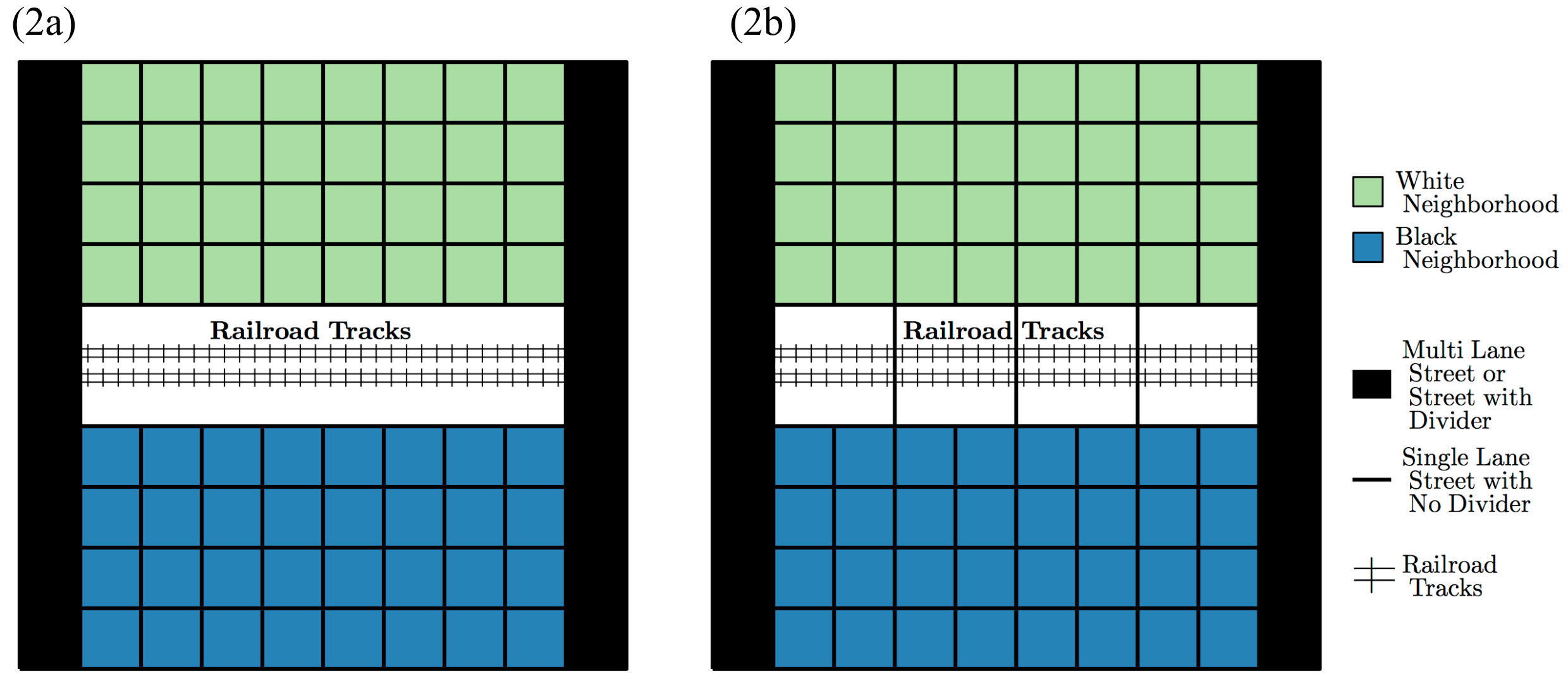

**Figure 3.** Example of the TIGER/Line Features used to Construct Road Networks

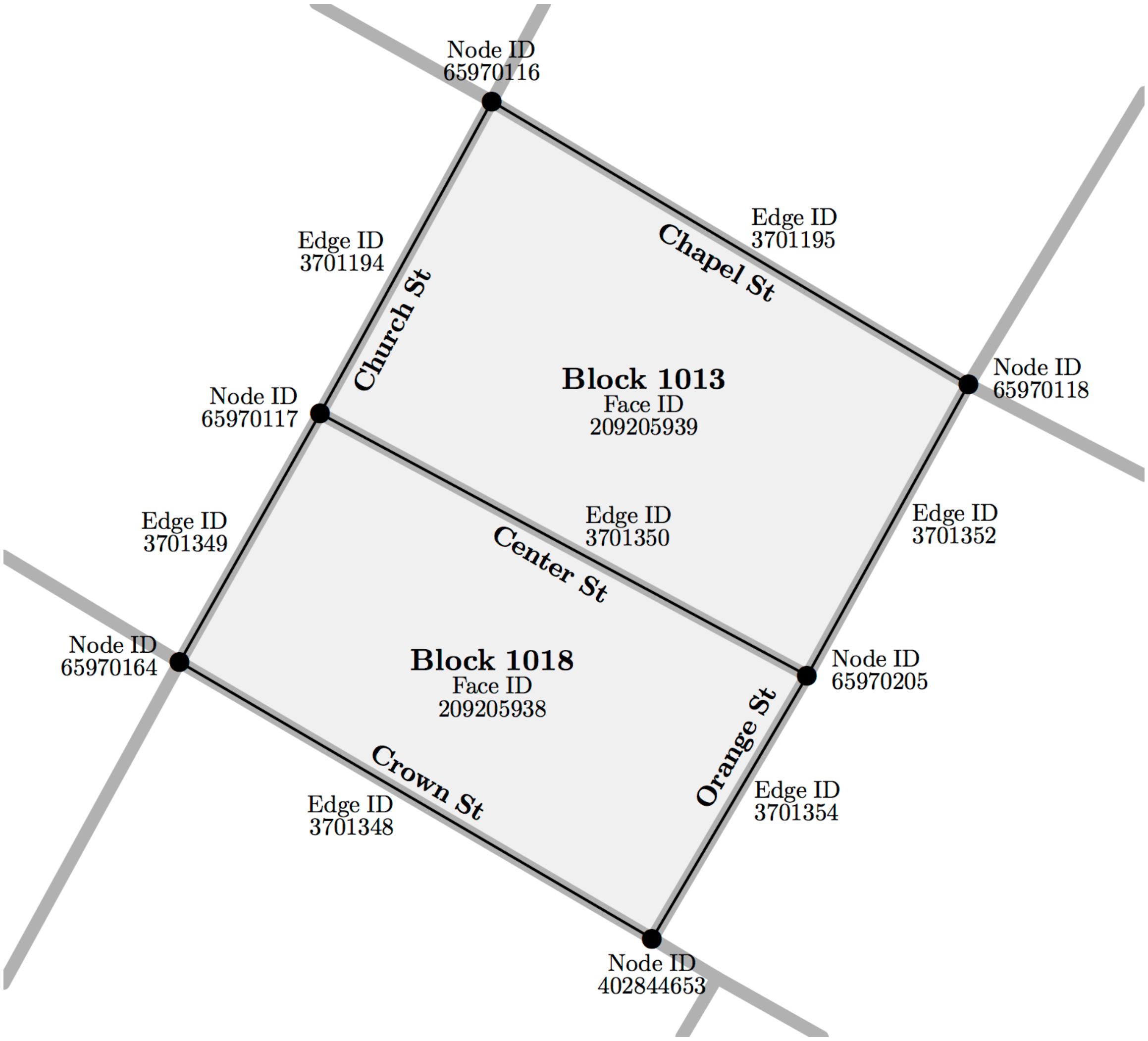

**Figure 4.** Example of How a Physical Barrier Affects Local Environments Using Straight Line versus Road Distance

4a.

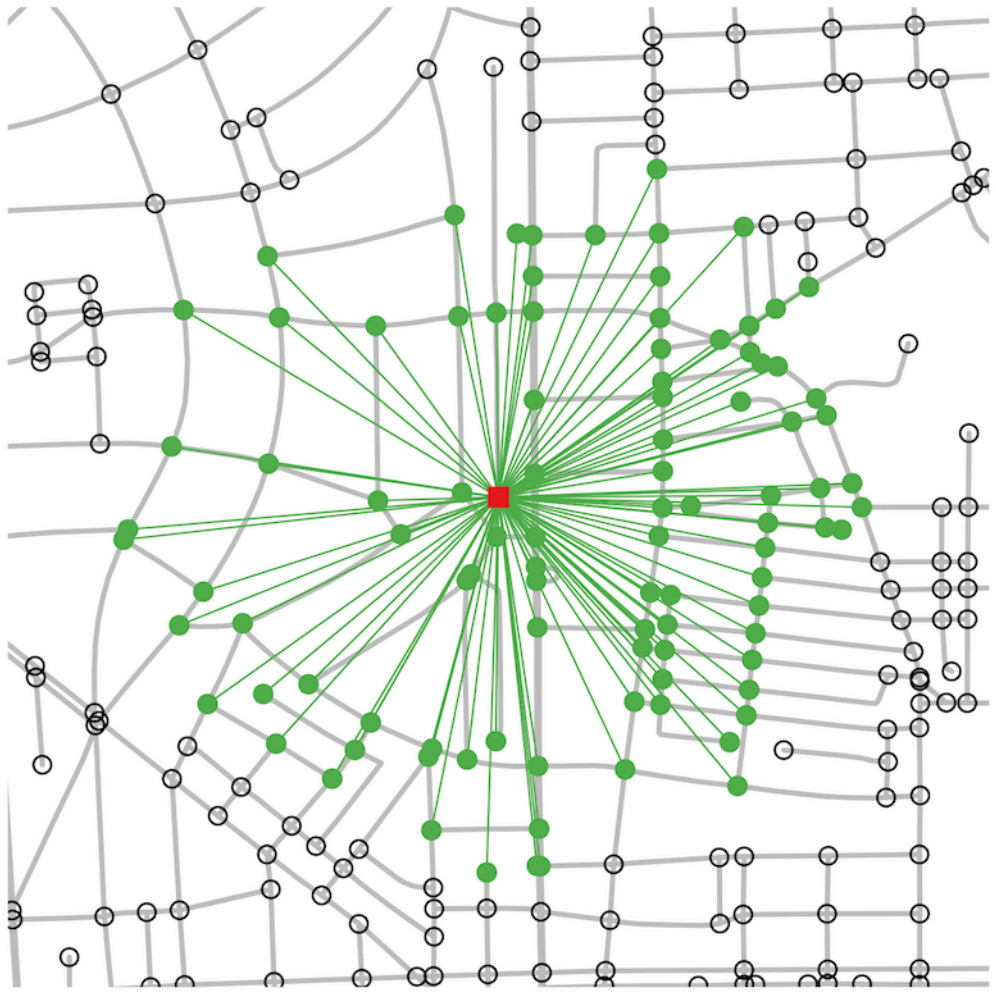

4b.

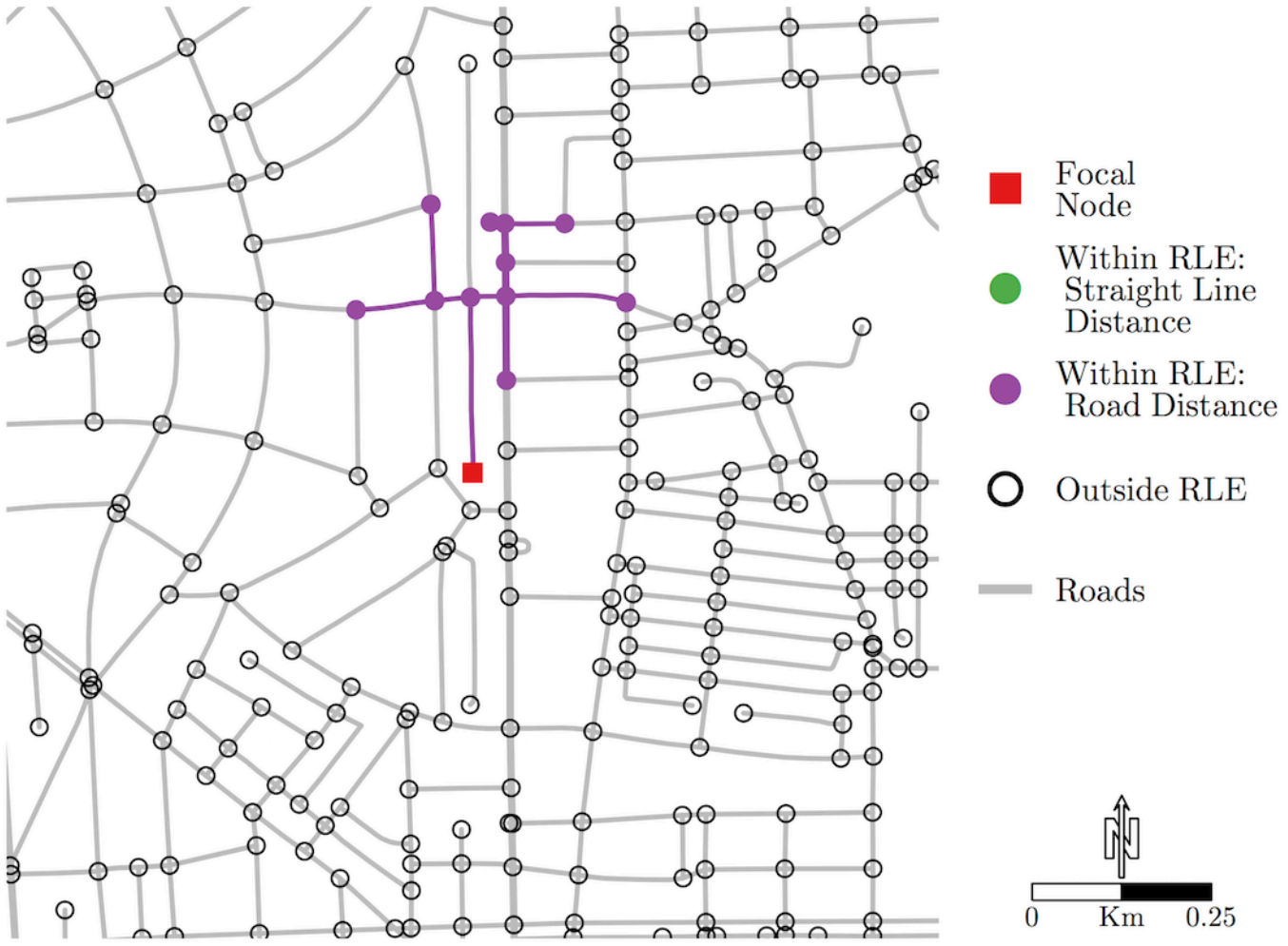

4c.

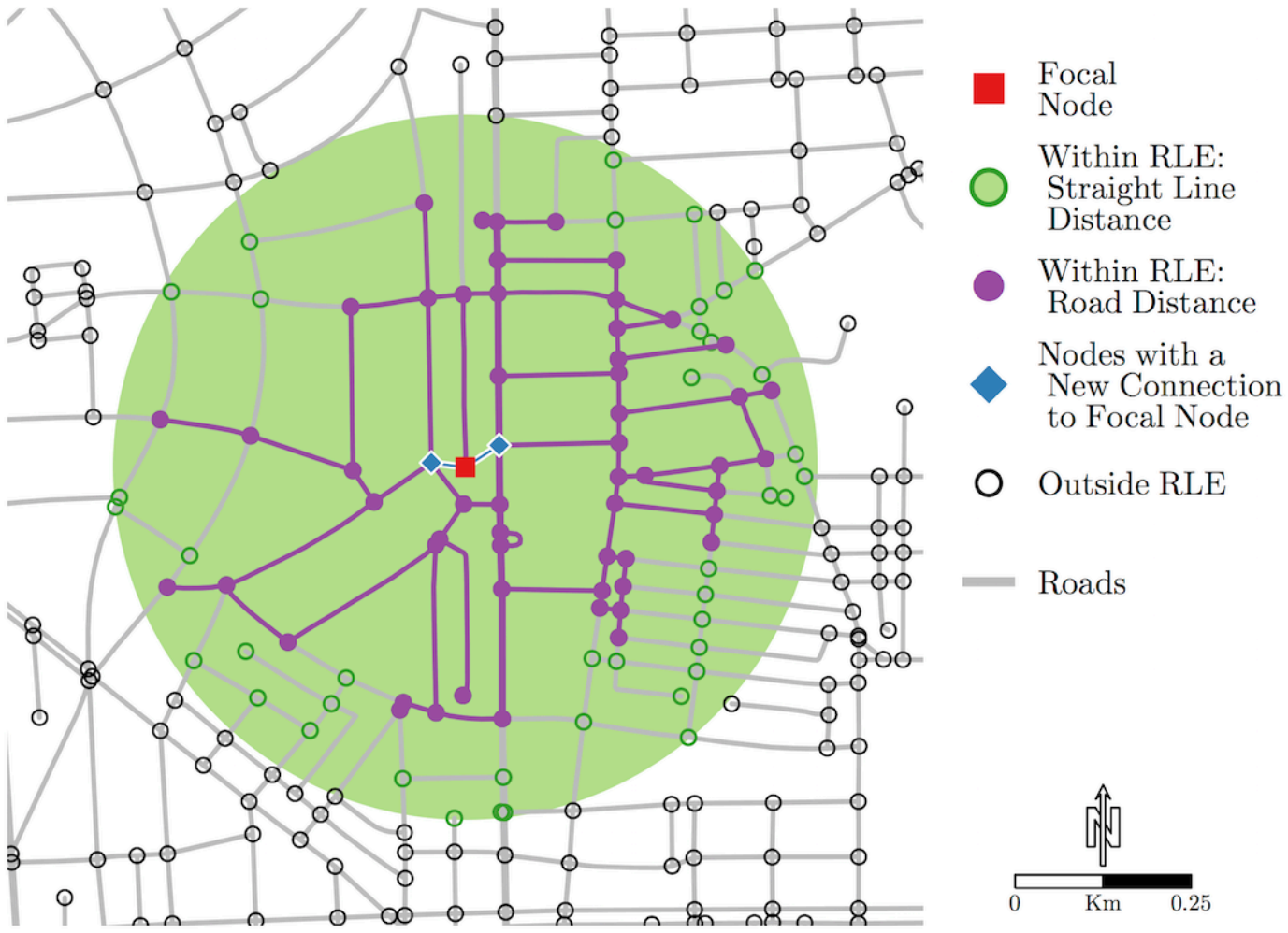

**Figure 5.** Segregation Indexes for Cities

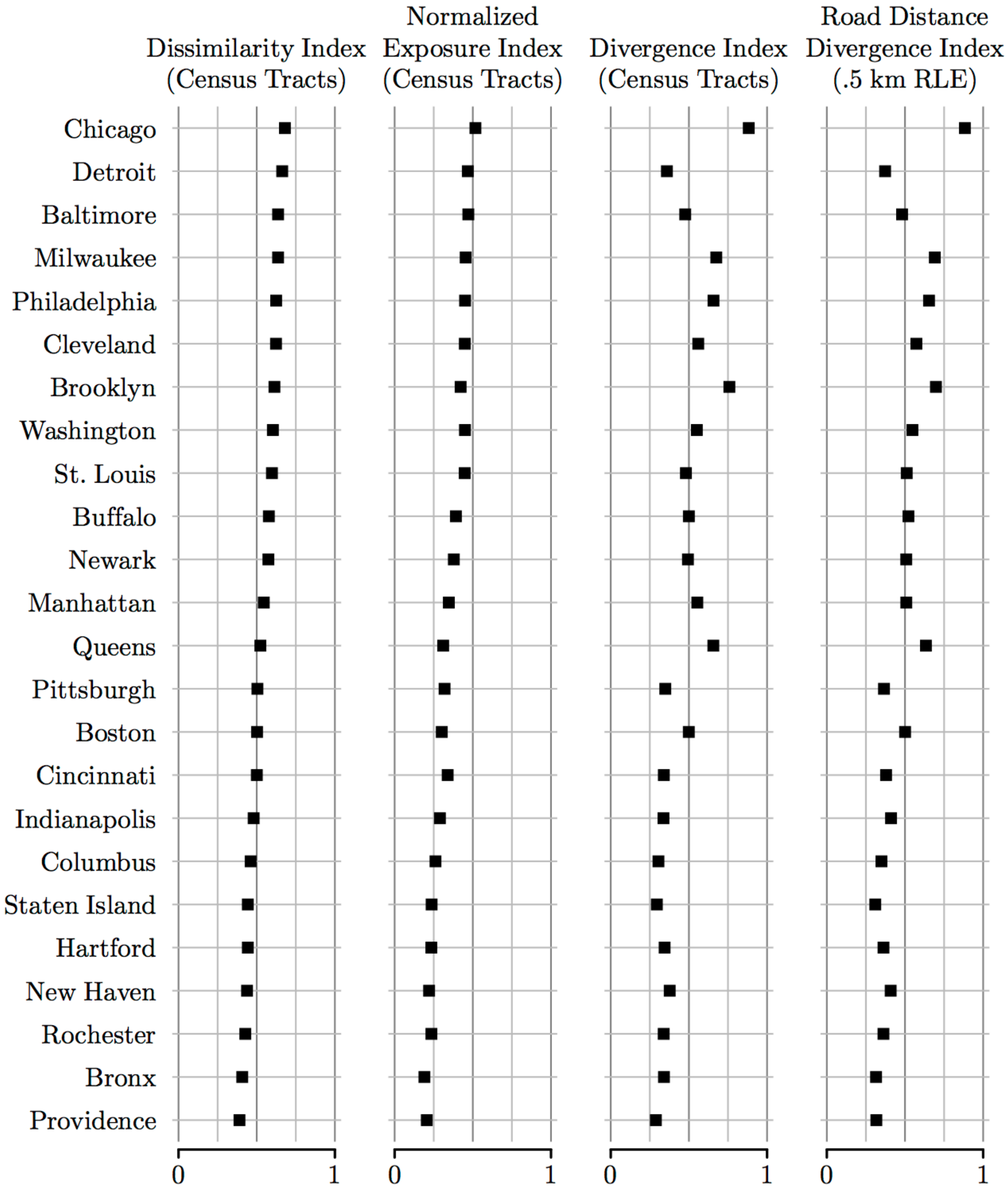

*Notes:* All segregation indices consider whites, blacks, Hispanics, and Asians using 2010 U.S. Census data.

**Figure 6.** Percent Difference between Road Distance and Straight Line Distance Divergence Indexes by City

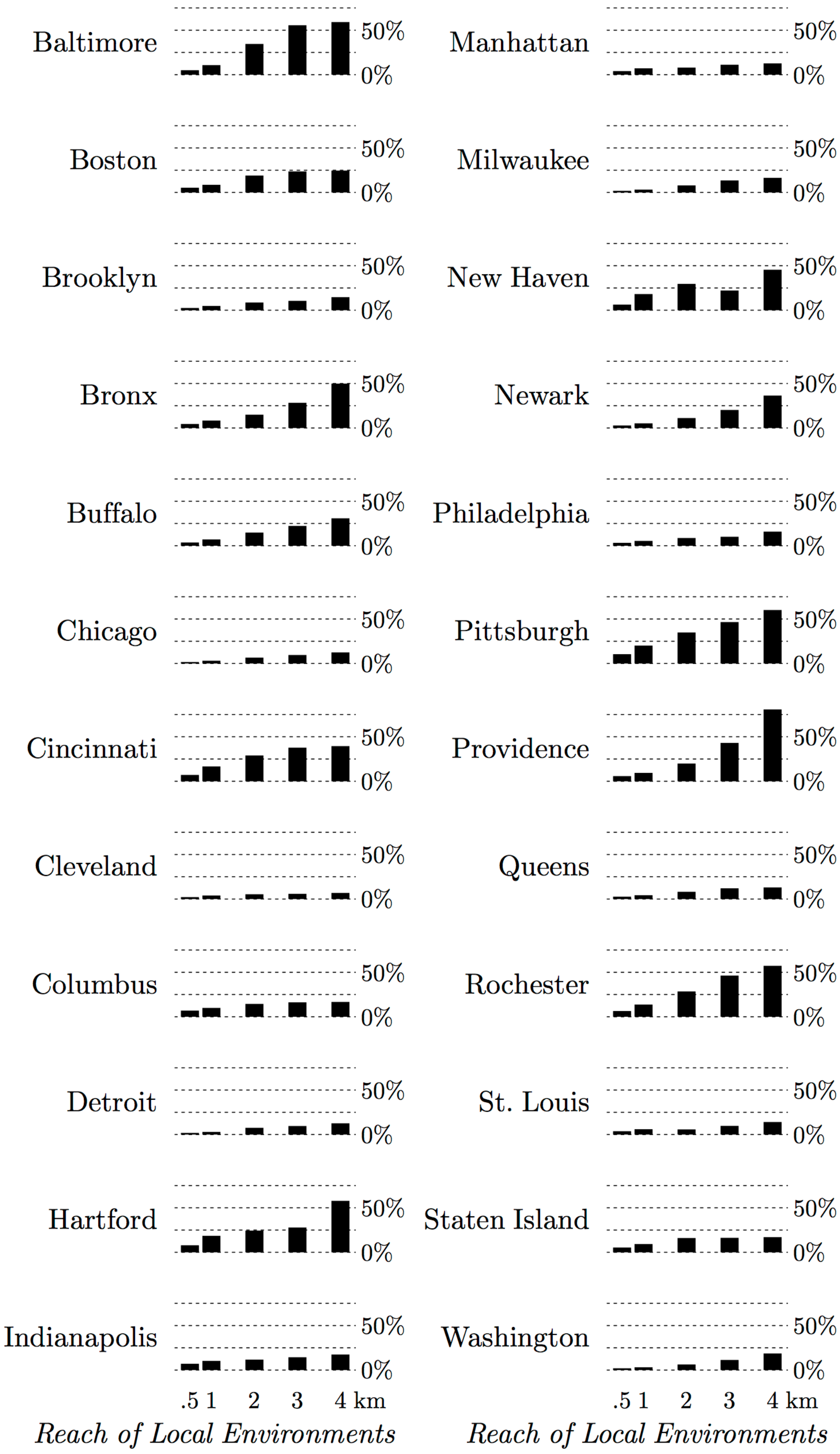

**Figure 7.** Quartiles of the Difference between Road Distance and Straight Line Distance Divergence Indexes in the Guilford and Waverly Neighborhoods of Baltimore using Local Environments with a Reach of .5 km

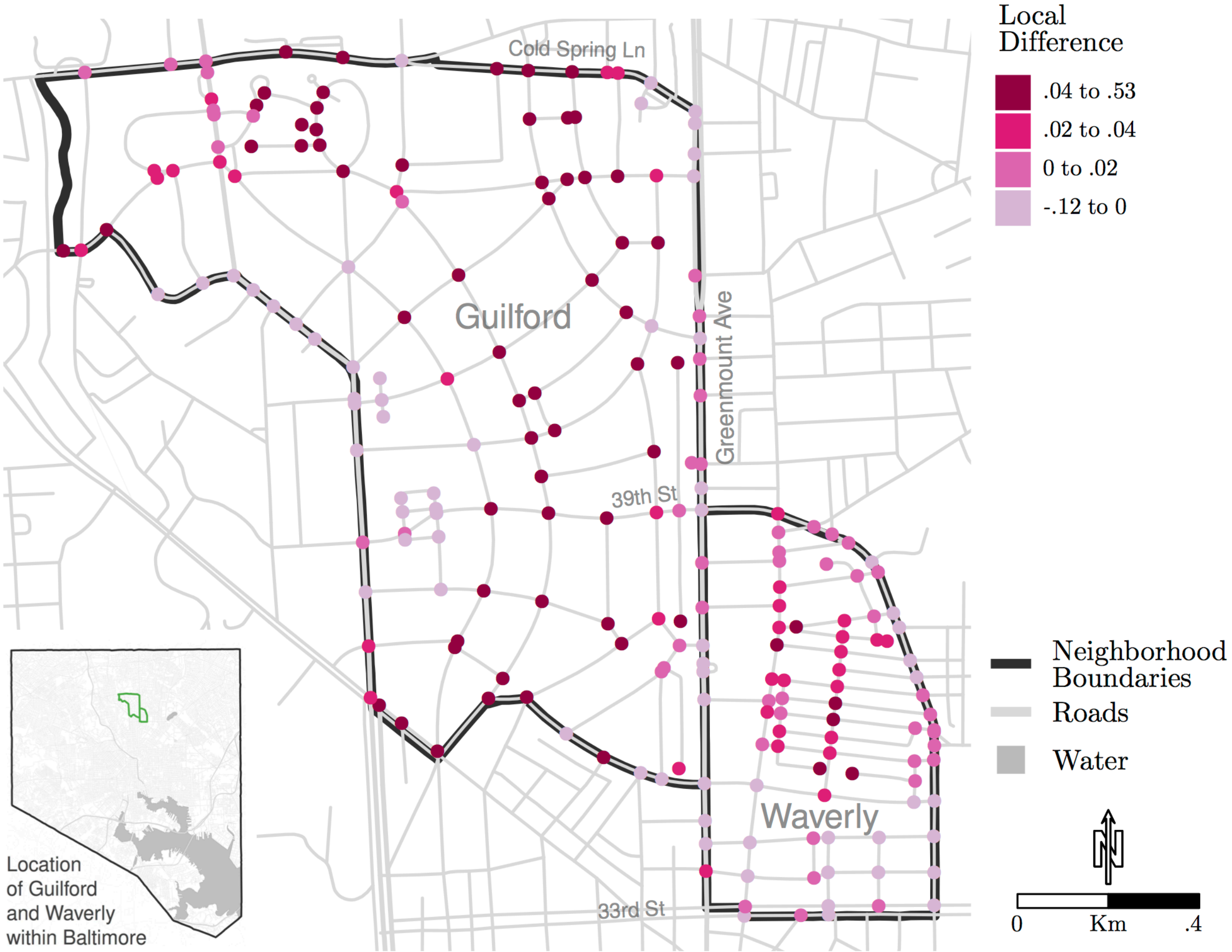

**Figure 8.** Difference between Road Distance and Straight Line Distance Divergence Indexes by Ethnoracial Group and City

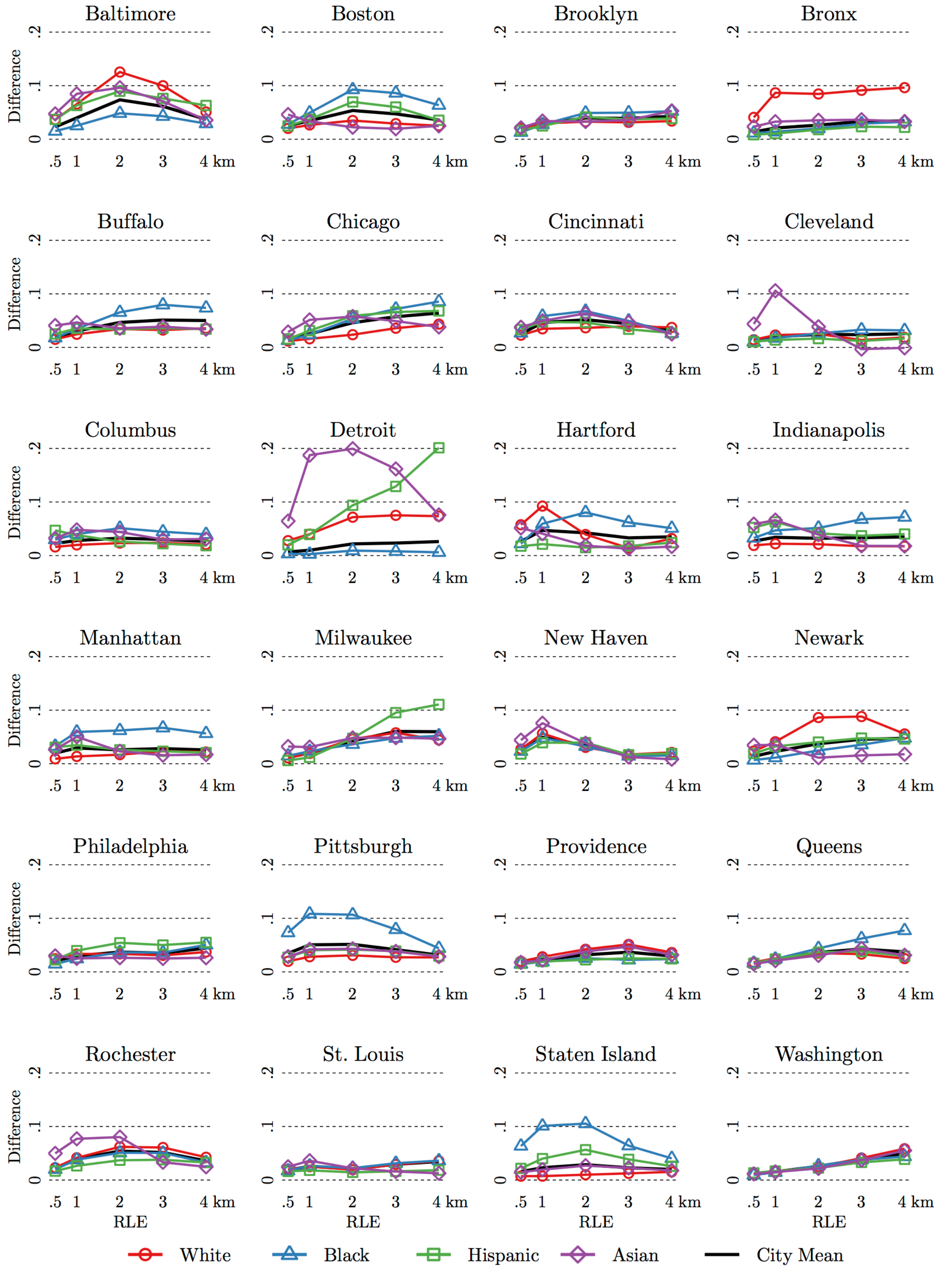

**Figure A1.** Example of How Adding Road Connectivity Increases the Number of Nodes in a Local Environment Constructed with Road Distance

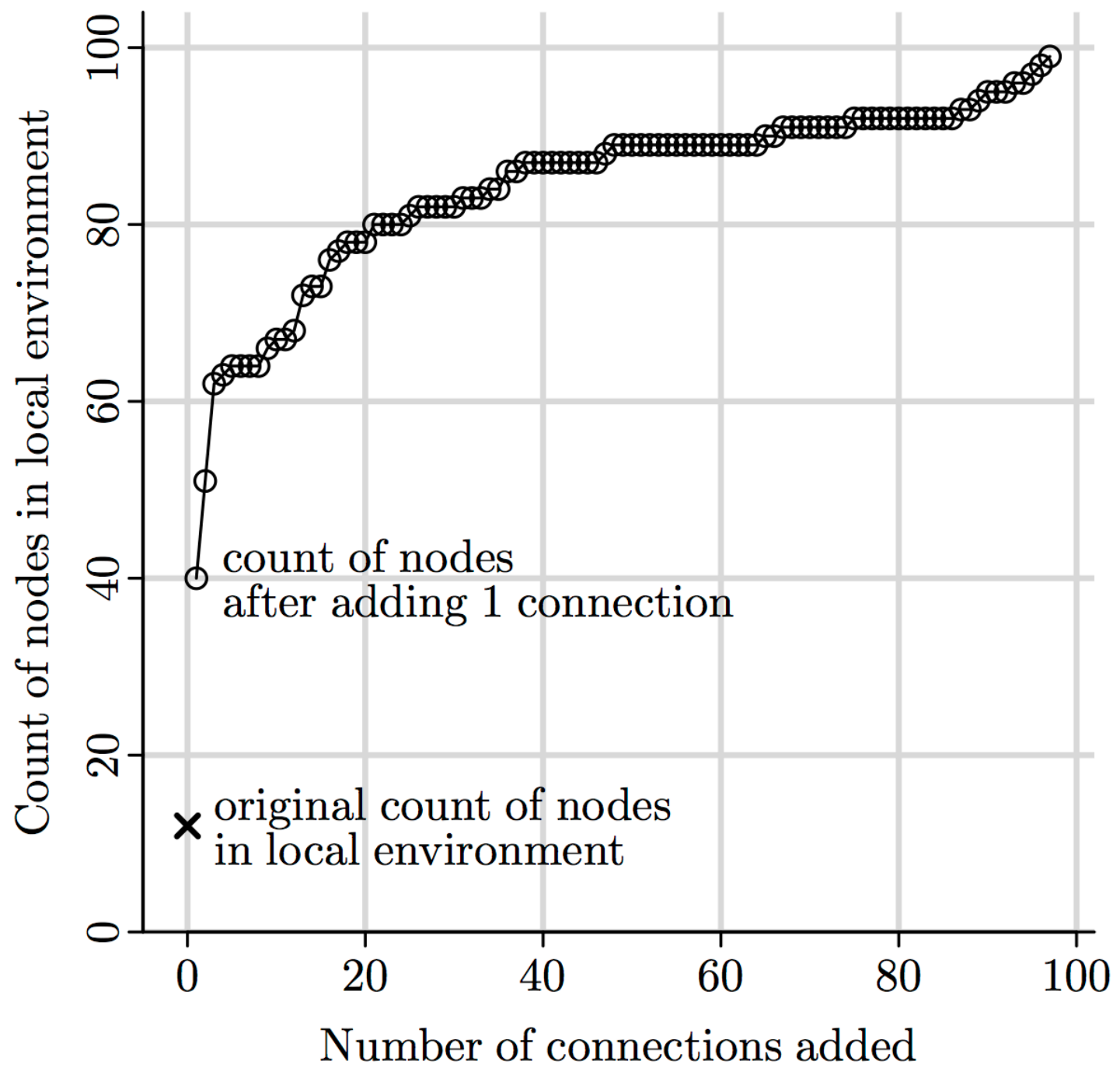

**Figure A2.** Difference between Road Distance and Straight Line Distance Divergence Indexes by City

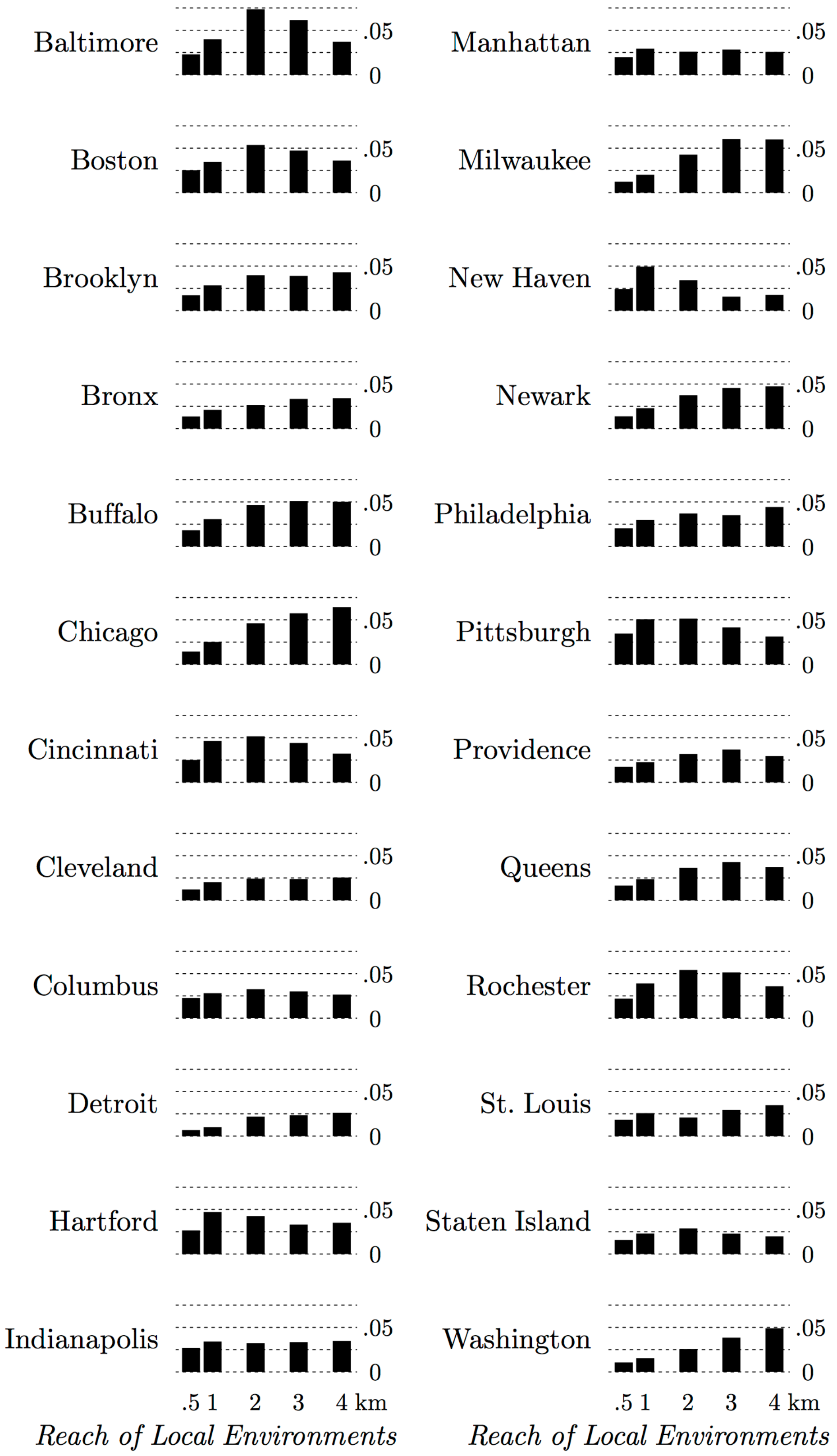

**Figure A3.** Percent Difference between Road Distance and Straight Line Distance Divergence Indexes by Ethnoracial Group and City

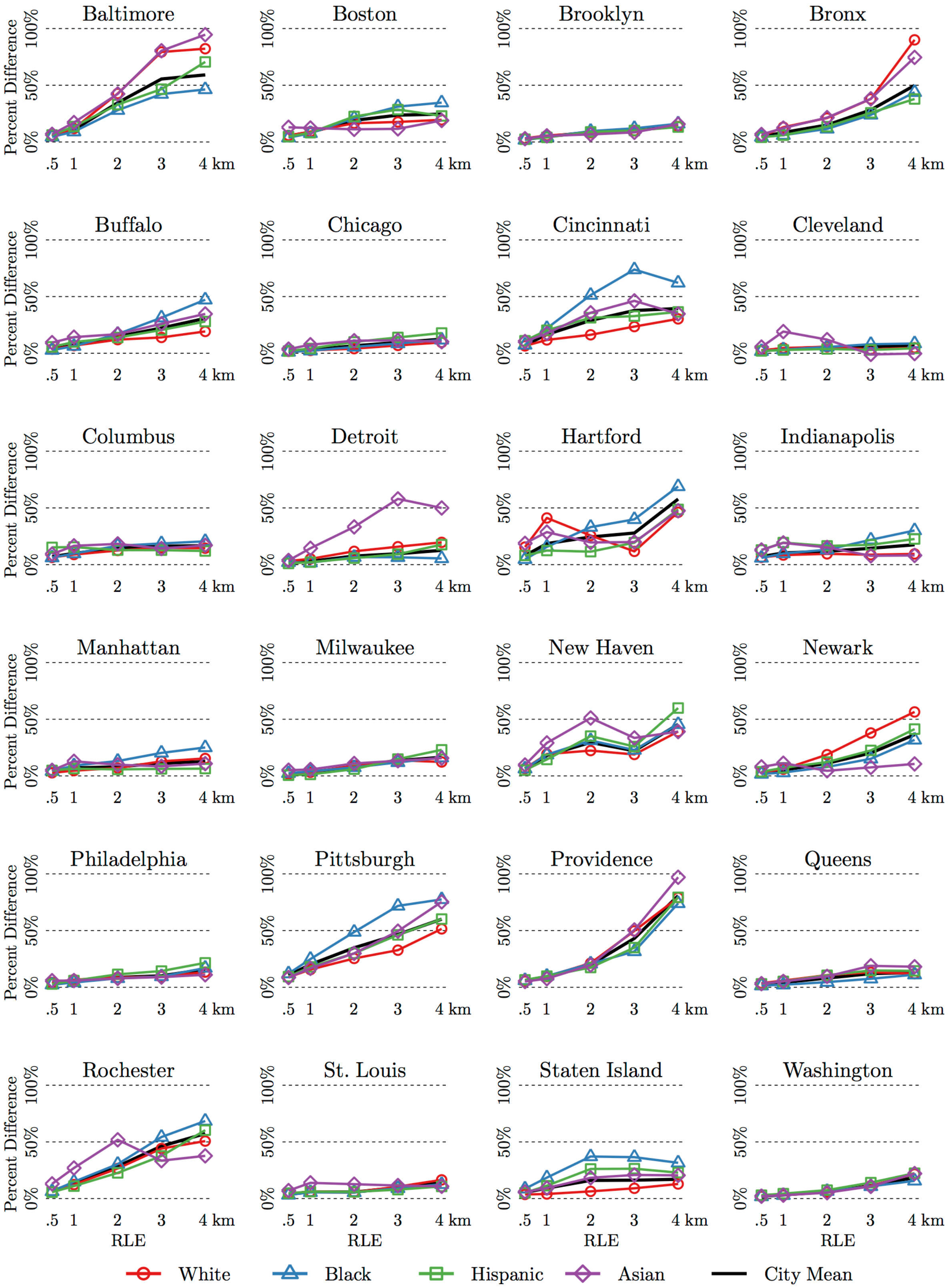

## APPENDIX

*Procedure for Assigning Block Populations to Road Locations*

We use the SPC method to distribute the aggregate population of each block to point locations on roads by assigning a portion of each block's population to the nodes associated with the block. The procedure allows us to calculate the population count and composition in the local environment around each node. We do this in two steps. First, we assign individuals to one of the roads associated with the block, with the probability of assignment equal to the length of the road segment. Second, we randomly assign individuals to one of the two nodes that are the end points of their assigned road segment, i.e. one of the intersections. When adjacent blocks are associated with the same node, the node will likely receive a portion of each block's population. It also has the advantage of removing the arbitrary administrative boundaries of individual census blocks, and it smoothes the distribution of the population and sharp discontinuities that may occur along the administrative boundaries.

The random assignment of block populations to nodes will affect the population count and composition of each node. The randomness of the procedure would likely affect segregation levels if we were measuring segregation aspatially and using each node as a unit of analysis. However, we measure segregation in the local environments around each node, which incorporates nearby populations into its composition. Even at a reach of 0 km, much of the variability of random assignment is mitigated, because adjacent blocks share intersections and each block contributes to the intersection's population. To err on the side of caution, we only examine results for local environments with a reach of at least .5 km, which is larger than a typical residential block in the cities we study, and variability in the population count or composition due to sampling is likely to be minimal.

*Measuring Segregation with the Divergence Index*

The Divergence Index for location $i's$ local environment with a reach of $r$ km is:

$$\widetilde{D}_{ri} = \sum_{m} \tilde{\pi}_{rim} \log \frac{\tilde{\pi}_{rim}}{\pi_m},$$

where $\pi_m$ is group $m$'s proportion in the overall population, and $\tilde{\pi}_{rim}$ is group $m$'s proportion of the proximity weighted population in location $i$'s local environment. The overall population is the city population.

The proximity weighted population composition for each location is calculated as:

$$\tilde{\pi}_{rim} = \frac{\int_{j \in K} \tau_{jm} \phi(i,j) dj}{\int_{j \in K} \tau_j \phi(i,j) dj},$$

where $\tau_j$ and $\tau_{jm}$ are the total and group-specific population counts for each location $j$ in region $K$, and $\phi(i,j)$ is the proximity function for locations $i$ and $j$. We use the following uniform proximity function to weight the influence of nearby and more distant locations:

$$\phi(i,j) = \begin{cases} 1 & \text{if } d(i,j) < r \\ 0 & \text{otherwise} \end{cases},$$

where $d(i,j)$ is the pairwise distance (straight line or road) between locations $i$ and $j$, and $r$ is the RLE.

A city's overall Divergence Index for a given RLE is the population weighted average of the Divergence Indexes for all locations, calculated as:

$$\widetilde{D}_r = \sum_i \frac{\tau_i}{T} \widetilde{D}_{ri} ,$$

where $T$ is the city population, and $\tau_i$ is the population of location $i$.

*Measuring Group-Specific Segregation*

To calculate group-specific segregation for each RLE, we calculate $\widetilde{D}_{ri}$ as above and then calculate the average degree of segregation experienced by each ethnoracial group as:

$$\widetilde{D}_{rm} = \sum_i \frac{\tau_{im}}{\tau_m} \widetilde{D}_{ri} ,$$

where $\tau_m$ is the population of group $m$ in the city, and $\tau_{im}$ is the population of group $m$ in location $i$. The weighted average of the group-specific segregation results is equal to the city's overall segregation:

$$\widetilde{D}_r = \sum_m \frac{\tau_m}{T} \widetilde{D}_{rm} .$$

[Table A1 about here.]

[Table A2 about here.]

[Table A3 about here.]

[Figure A1 about here.]

[Figure A2 about here.]

[Figure A3 about here.]